\documentclass[preprint]{revtex4-2}
\usepackage{hyperref}
\usepackage{graphics}
\usepackage{latexsym,exscale,stmaryrd,amssymb,amsmath}
\usepackage{tikz}
\usetikzlibrary{arrows}
\usetikzlibrary{positioning}
\usetikzlibrary{fit}
\usepackage{subfigure}
\usepackage{tabularx}
\usepackage{booktabs}
\usepackage{multirow,bigstrut}
\begin{document}
\pagenumbering{arabic}
\title{Coarse graining of biochemical systems described by discrete stochastic dynamics}
\date{\today} 

\author{David Seiferth$^{(1)}$}
\author{Peter Sollich$^{(2)}$}
\author{Stefan Klumpp$^{(1)}$}
\affiliation{$^{(1)}$ Institute for the Dynamics of Complex Systems, University of G\"ottingen, Friedrich-Hund-Platz 1, 37077 Göttingen, Germany}
\affiliation{$^{(2)}$ Institute for Theoretical Physics, University of Göttingen, Friedrich-Hund-Platz 1, 37077 Göttingen, Germany}

\begin{abstract}
Many biological systems can be described by finite Markov models. A general method for simplifying master equations is presented that is based on merging adjacent states. The approach preserves the steady-state probability distribution and all steady-state fluxes except the one between the merged states. Different levels of coarse graining of the underlying microscopic dynamics can be obtained by iteration, with the result being independent of the order in which states are merged. A criterion for the optimal level of coarse graining or resolution of the process is proposed, via a trade-off between the simplicity of the coarse-grained model and the information loss relative to the original model. As a case study, the method is applied to the cycle kinetics of the molecular motor kinesin.
\end{abstract}

\maketitle

\section{\label{sec:intro}Introduction}
Many stochastic systems in physics, chemistry, biology as well as other areas can be described by
discrete-state Markov models, such that their dynamics is given by a master equation \cite{schnakenberg, VANKAMPEN200796}. Examples include the dynamics of (bio-)molecules where states describe discrete configurations or functional states, the dynamics of chemical reactions and of populations, where the states are given by the numbers of individuals of each species, discrete stepping of molecular machines, and many more \cite{mcquarrie_1967, Gillespie_2007}. In all these cases the dynamics can be interpreted and visualized as a hopping process on a network where nodes represent the states of the system and links between them transitions from one state to another \cite{schnakenberg,hill}.

Often, these discrete models can be understood as arising from more complex underlying dynamics by coarse graining. For example in the case of macromolecules, the dynamics of the discrete configurations of the molecules represent, on the microscopic level, the motion of the atoms in that molecule \cite{Noe19011}. Coarse graining, the operation that transforms the microscopic description into the simpler macroscopic one, thus involves the elimination of degrees of freedom and a reduction of the dimension of the state space. Coarse graining can make large systems computationally tractable, but can also have conceptual advantages over the more detailed models by focusing on the essential parts of the dynamics.

In general, coarse graining can be applied to systems with discrete or continuous degrees of freedom and coarse-grained models may also be of either type. For example in molecular dynamics, coarse-grained models of macromolecules are often obtained by a united atom approach \cite{Saunders_Voth,Praprotnik_Kremer}, i.e.\,by merging groups of atoms into a single larger unit, which may be spherical \cite{Weiner1984} or anisotropic \cite{Anisotropic_potential}. The larger unit has fewer (but still continuous) degrees of freedom than the original atoms.  As already mentioned, the case of coarse graining a continuous description into a simpler discrete one is also  important for describing the molecular dynamics of macromolecules \cite{Noe19011, Noe_2011, SCHUTTE1999146}. 
Another case of interest is the simplification of a given discrete-state system, as very often descriptions of the same system at different levels of detail are possible. An example is given by the chemomechanical dynamics of a molecular machine, which moves through a working cycle in multiple steps. Depending on what a theoretical description of that system is used for, fewer or more of these steps need to be explicitly included in the model \cite{Klumpp_Motors}. Likewise, alternative cycles and events branching off from the main cycle may or may not be part of such description. This can be done ad hoc, but to simplify descriptions in a controlled manner, systematic protocols for coarse graining are needed. 

Several coarse-graining approaches for discrete stochastic systems have recently been proposed and studied \cite{Hummer,SlowFastStates,Vollmer-CG}. They all have advantages and disadvantages, which we will discuss in detail below. In any case, coarse graining is accompanied by a loss of information due to the simplification of the original system. Therefore most coarse-graining approaches attempt to simplify a system while retaining as much as possible its key characteristics. Important challenges in this process are to identify microstates that can be faithfully coarse grained and to determine which degree of coarse graining is optimal in some sense.
Here, we present an iterative approach that in each step merges two adjacent states in a finite discrete-state Markov model and results in a hierarchy of models with different levels of coarse graining.
Our approach does not impose any constraints on the underlying network topology and hence can be used to merge states in any system governed by a master equation.
By merging states, the dimensionality of the state space decreases, the system becomes simpler and some information is lost. In our approach specifically, the transition flux between the merged states vanishes and the dynamics of the system can be modified. However,  
transition rates and steady-state probabilities are only changed locally and transition fluxes are retained.
The steady-state distribution and the associated probability fluxes were proposed to provide a complete and unique description for any steady state \cite{Zia_2006}. Therefore, we impose only local changes in these quantities to approximate the original model well. 
Since our coarse-graining procedure generates a hierarchy of models, we discuss a cost function that relates the information loss and the gain in simplicity. This cost function can be used to find an optimal balance between information loss and simplicity and, thus, an optimised degree of coarse graining.

The paper is organised as follows:
After introducing the master equation and related quantities in section \ref{sec:theoretical_approach}, we present our coarse-graining procedure and show how transition rates in the coarse-grained system can be defined without changing the flux between any pair of states. An alternative derivation by minimizing the Kullback-Leibler divergence for trajectories provides further motivation for the choice of transition rates in the coarse-grained system. A detailed comparison of different coarse-graining approaches is also included in this section. As a case study of a prototypical system in a non-equilibrium steady state, we apply our approach to the molecular motor kinesin and its multi-cyclic kinetic diagram as proposed in ref. \cite{Liepelt}. First, we study coarse graining of that kinetic diagram without changing its cycle topology in section \ref{sec:Kinesin}. Then, in section \ref{sec:Changed_cycle_typology}, we generalize the approach and also allow changes in the number of cycles. This allows us to derive uni-cyclic models, which have originally been used to describe the kinesin motor \cite{Fisher}, from the multi-cyclic model for a systematic comparison. 
Furthermore, we present a criterion that balances simplicity and information loss compared to the original model to find the optimal level of coarse graining from the hierarchy of models that are coarse-grained to different degrees. 

\section{Coarse-graining framework\label{sec:theoretical_approach}}
\subsection{Master equation and cycle decomposition}
We consider discrete stochastic systems described by a master equation. The system has $N$ discrete states and transition between states $i$ and $j$ occur with time-independent transition rates $\alpha_{ij}$. At a certain time $t$,  a state $i$ is occupied with probability $p_i(t)$, which evolves according to the master equation 
\begin{equation}
\dfrac{d p_i}{dt}=\sum_{j\neq i} (\alpha_{ji}p_j - \alpha_{ij}p_i)=\sum_{j\neq i} J_{ji}.
\label{eq:ME}
\end{equation}
The master equation can be interpreted as balance of incoming ($\alpha_{ji}p_j $) and outgoing fluxes ($\alpha_{ij}p_i$).
In a steady-state, the occupation probabilities are time-independent and hence the left-hand side of the equation is zero. In the following, we will focus on steady-state quantities unless stated otherwise. Steady-state probabilities can be calculated by matrix inversion of the master equation or in terms of a sum over spanning trees of a graph representing the states and transitions  \cite{hill}.
In equilibrium, there is no probability flux between any pair of states, a condition called detailed balance, as
\begin{equation}
\alpha_{ij}p_i=\alpha_{ji}p_j 
\label{DB}
\end{equation}
for all pairs of states $i$ and $j$. Here, we will consider systems in non-equilibrium steady-states. Hence, while the occupation probabilities are time-independent, $\partial_t p_i=0$, the steady-state flux between a pair of states is in general not zero. The net flux from state $i$ to $j$ measures the net number of transitions per time and is given by
\begin{equation}
J_{ij}=\alpha_{ij}p_i-\alpha_{ji}p_j.
\end{equation}
All nonzero fluxes are associated with the production of entropy, a feature that we will use below 
to quantify the information loss between a fine-grained and a coarse-grained description of the same system. 
The total entropy production $P$ can be expressed in terms of the net-transition fluxes $J_{ij}$ and the corresponding transition affinities $\Delta S_{ij}$ \cite{schnakenberg}, 
\begin{equation}
P=\dfrac{1}{2}\sum_{i,j} J_{ij} \Delta S_{ij}.
\end{equation}
The transition affinity 
\begin{equation}
\Delta S_{ij}=\ln  \dfrac{\alpha_{ij}p_i}{\alpha_{ji}p_j}
\end{equation}
quantifies the change of entropy associated with a transition. 
The entropy production can be used to quantify the deviation from detailed balance for non-equilibrium systems \cite{Szabo_EntropyProd}. 

\begin{figure}
\centering
\includegraphics[scale=0.9]{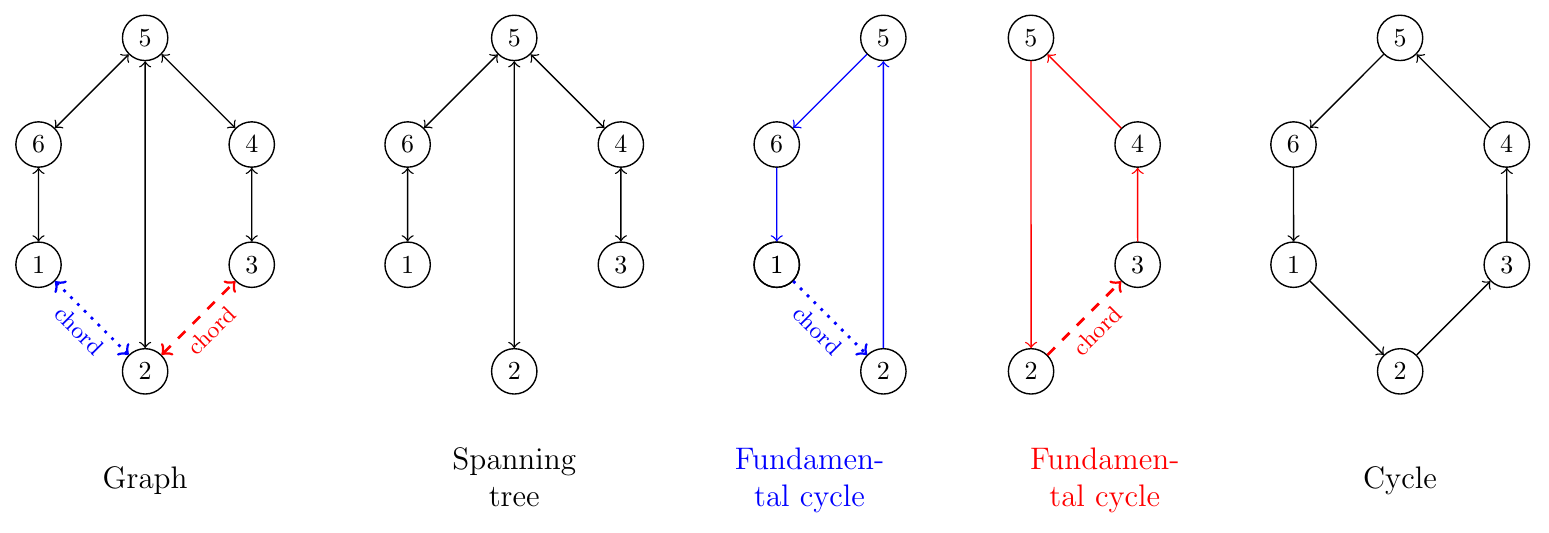}
\caption{A graph with one arbitrary chosen spanning tree. The edges missing in the spanning tree are called chords and define fundamental cycles. The blue dotted edge $1\rightarrow 2$ is a chord and defines the first (fundamental) cycle, while the red dashed chord $3\rightarrow 4$ defines the second (fundamental) cycle. The third cycle can be constructed by fundamental cycles \cite{schnakenberg}. The choice of a spanning tree of a graph is arbitrary. 
The reference orientation of all cycle is chosen to be counterclockwise.
\label{fig:Cycle_Decomposition} }
\end{figure}
For many purposes, it is useful to represent states and transitions as the nodes and edges of a graph. Doing so provides a constructive method for obtaining the steady state probabilities \cite{schnakenberg,hill}. Moreover, it allows us to decompose the dynamics on the network into a set of fundamental cycles and to express fluxes as well as the entropy production as sums over these fundamental cycles \cite{schnakenberg}. 

All nonzero fluxes in the steady state are associated with cycles, closed paths of nodes and edges in the graph. A large graph will typically contain multiple cycles, of which not all are independent of each other. In fact, all cycles can be constructed from a set of fundamental cycles, which in turn can be constructed from spanning trees. A spanning tree is a connected subgraph that contains all nodes, i.e. all states of the system, but no cycles. It is obtained from the full graph by removing $L$ edges.
Adding back one  edge to such a tree yields a single cycle.
The added edge is called chord $l$ and defines a corresponding fundamental cycle $C_l$. Figure \ref{fig:Cycle_Decomposition} shows a 6-state network with three cycles of which two are fundamental. The choice of a spanning tree is arbitrary. Different spanning trees can result in different sets of fundamental cycles.
The number of fundamental cycles is defined by the number $L$ of chords of a network. In a network with $E$ edges and $N$ nodes, there are
\begin{align}
    L=E-N+1
    \label{eq:Number_chords}
\end{align}
chords \cite{schnakenberg}.
Moreover, all transition fluxes can be expressed as a linear combination of chord fluxes \cite{schnakenberg}. Hence, a network with $L$ chords has $L$ independent transition fluxes and $L$ fundamental cycles. There are two chords in figure $\ref{fig:Cycle_Decomposition}$ and hence two fundamental cycles.  

The entropy production can then be expressed in terms of quantities related to fundamental cycles or to the corresponding chords $l$ as
\begin{align}
P=\sum_l J_l \Delta S_{C_l}, 
\label{eq:entropy_prod_decomposition}
\end{align}
where the sum runs over a set of fundamental cycles. In this expression, $J_l$ is the chord flux, the net-number of transitions between states connected by the chord $l$ per unit time. $\Delta S_{C_l}$ is the affinity of the fundamental cycle $C_l$ defined by chord $l$, that is the entropy change 
after one completion of the cycle. The cycle affinity can be written as
\begin{align}
\Delta S_{C_l} =\sum_{(ij)\in C_l} \Delta S_{ij} = \ln \left( \prod_{(ij) \in C_l} \dfrac{\alpha_{ij}}{\alpha_{ji}} \right),
\label{eq:Cycle_Affinity}
\end{align} 
where both sum and product are evaluated over all edges $(ij)$ within cycle $C_l$. We choose an arbitrary reference orientation for each fundamental cycle. In figure \ref{fig:Cycle_Decomposition}, the reference orientation is chosen to be counterclockwise.

\subsection{Coarse-graining procedure \label{sec:CG_procedure}}
\begin{figure}
    \centering
    \includegraphics{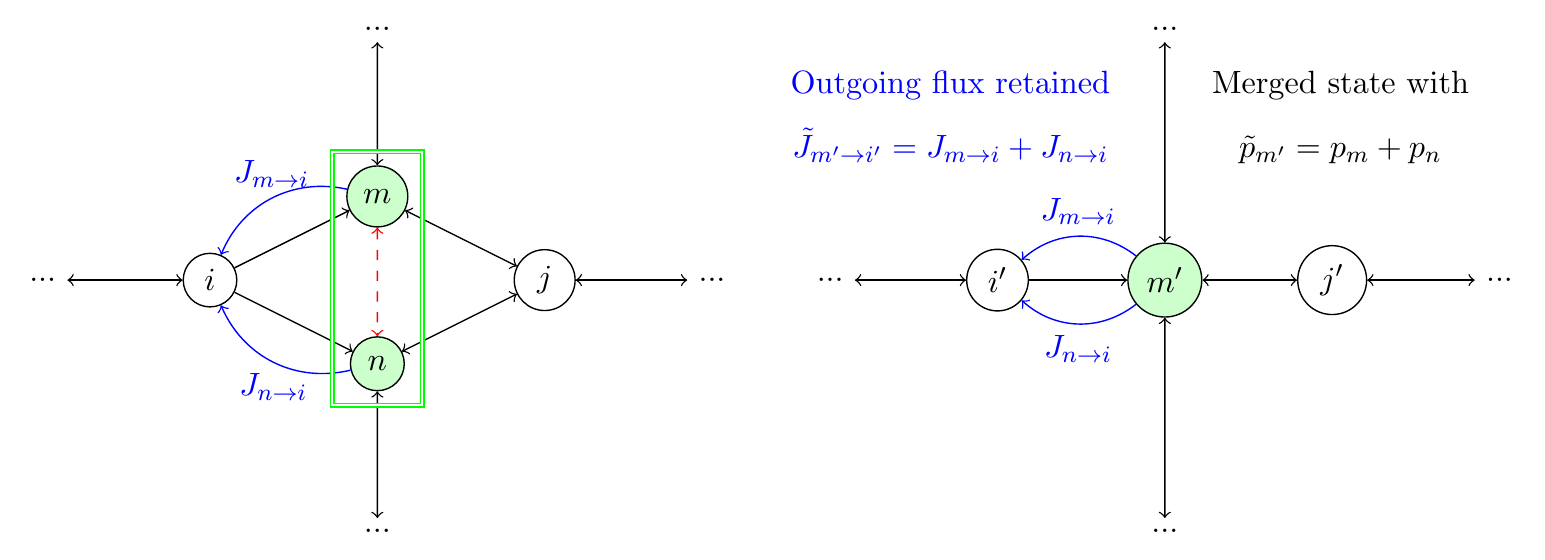}
    \caption{State $m$ and state $n$ will be merged such that the fluxes between the merged state and the rest of the system are retained and the probabilities add up.\label{fig:Bsp_Coarse_graining}}
\end{figure}
In this section, we present a procedure for coarse graining discrete stochastic systems. To simplify large networks, we merge two connected nodes $m$ and $n$. 
Further reduction can be achieved by iterating this procedure, as we will discuss below.  In general, the dynamics of the reduced network with two nodes merged is non-Markovian. However, we approximate the dynamics with a modified master equation. The error of this approximation will be smallest if there is a time scale separation (discussed in section \ref{sec:comparison}). Thus, we require that (approximately) the coarse-grained system shall still obey a master equation
\begin{align}
\dfrac{d \tilde{p}_i}{dt}=\sum_{j\neq i} (\tilde{\alpha}_{ji}\tilde{p}_j - \tilde{\alpha}_{ij}\tilde{p}_i)
\label{eq:ME_cg}
\end{align}
but with modified probabilities and transition rates. We adopt the following notation: As before, we denote the original probabilities and rates by $p_i$ and $\alpha_{ij}$. The probabilities of states in the coarse-grained system are denoted by $\tilde{p}_i$ and the transition rates by $\tilde{\alpha}_{ij}$. 
Note that in the following we derive constraints that refer only to steady-state probabilities.
To conserve the probabilities, we request that the steady state probabilities of the two merged states add up, while all others are unaffected,
\begin{align}
\tilde{p}_i=
\begin{cases}
p_i &\text{ for }i\neq m, i\neq n
\\
p_m+p_n &\text{ for }i=m.
\end{cases}
\label{eq:probailities_cg}
\end{align}
To determine the rates of the coarse-grained system, we impose the following additional requirements.
\begin{enumerate}
\item All transition rates that do not point to or from one of the two merged nodes remain unchanged, $\tilde{\alpha}_{ij}=\alpha_{ij}$ for $i,j\neq m,n$.
\item All one-way fluxes flowing from or to a merged node stay the same: outgoing (one-way) fluxes from the merged nodes $n$ and $m$ to the same node $i$ are added, $\tilde{J}_{m\rightarrow i}=J_{m\rightarrow i}+J_{n\rightarrow i}$. Likewise, incoming (one-way) fluxes to the merged nodes are added as well, $\tilde{J}_{i\rightarrow m}=J_{i\rightarrow m}+J_{i\rightarrow n}$. 
\end{enumerate}
Since the probabilities of nodes other than $n$ and $m$ stay the same, the latter condition directly implies that the incoming rate in the coarse-grained network is $\tilde{\alpha}_{i m}=\alpha_{i m}+\alpha_{i n}$. 
The condition on outgoing fluxes determines the rates of transitions from the merged node to any other node. 
Thus, all requirements stated above are fulfilled if the transition rates $\tilde{\alpha}_{ij}$ of the coarse-grained system are defined by
\begin{equation}
\begin{aligned}
\tilde{\alpha}_{ij}&=\alpha_{ij} & \text{ for } i,j \neq m,n \\
\tilde{\alpha}_{mj}&=\dfrac{\alpha_{mj}p_m+\alpha_{nj}p_n}{p_m+p_n}\\
\tilde{\alpha}_{im}&=\alpha_{im}+\alpha_{in} ,
\label{eq:general_transition-rates}
\end{aligned}
\end{equation}
where states $m$ and $n$ have been merged and states $i\neq m, n$ and $j \neq m,n$ are arbitrary states.

In this way, only transition rates pointing to or from the merged nodes are changed. The steady-state probabilities of the states that have not been merged stay the same and add up for the merged state. \textcolor{blue}{}The transitions going into a merged state are a sum. The transition coming out of a merged state are a weighted sum of the transitions in the original system. All fluxes are retained except the one between the merged states.
Other quantities may be different in the coarse-grained system. These include the net-cycle fluxes, i.e. the number of cycle completions per time, the corresponding thermodynamic forces, and the energy dissipation. 

The procedure of merging two adjacent states as described above can be iterated to further reduce the size of the system. Importantly,  the result is independent of the order in which states are merged.
We can differentiate two cases: On the one hand, we can merge first one pair of states and then another one, that has no direct transition to the first pair. Our approach incorporates only local changes and hence both coarse-graining iterations are independent. On the other hand, we can merge first a pair of states and then another one that is directly linked to the first pair. In that case, we can check by iterating equation \eqref{eq:general_transition-rates} that merging states $m$ and $n$ with a third state $k$ is associative. 

As the described approach can be applied iteratively and the order of the steps does not affect the results, we can use it to obtain a hierarchy of different coarse-grained models of the same underlying microscopic dynamic.
In general, we can define a mapping such that several states $i$ in the original system are lumped together in one state $I$. States in the original system are denoted by small letters and states in the coarse-grained system by capital letters. For a given mapping, the rates in the coarse-grained system are then given by
\begin{align}
\tilde{\alpha}_{IJ}=\dfrac{\sum_{i \in I, j \in J}\alpha_{ij}p_i}{\sum_{i\in I} p_i}.
\end{align}

Here we have derived the probabilities and rates of the coarse-grained system from a set of plausible requirements imposed to make the steady-state properties of the simplified system similar to the ones of the original system. In section \ref{sec:KL}, we will give an alternative derivation and motivate the choice of transitions rates in the coarse-grained system by minimizing the Kullback-Leibler divergence for trajectories.

Our coarse-graining approach  requires the calculation of the steady-state distribution of the original system, which can be computationally costly for large systems. This requirement obviously limits the usefulness of our method in practice, in particular, when it is applied to large scale systems. However, the approach allows us to study the dynamics of the system exclusively at the coarse-grained level (without need to solve the microscopic dynamics) in an approximate fashion. In addition, it provides a systematic way to compare different degrees of coarse-graining and to find the optimal coarse-graining level, as we will discuss below.

\subsection{Coarse-grained energy landscape in equilibrium systems\label{sec:cg_energy_landscape}}
We briefly consider the special case of canonical equilibrium systems.
The probability distribution for systems in equilibrium is given by the Boltzmann distribution
\begin{align}
p_i = \dfrac{1}{Z}e^{-\beta E_i},
\label{eq:Boltzmann}
\end{align}  
where $Z$ is the partition sum, $\beta = 1/k_B T$ is the inverse of Boltzmann's constant and thermodynamic temperature and $E_i$ the energy of state $i$. We compare the energy landscape of the original and coarse-grained system to make a consistency check of our approach. If we merge state $m$ and state $n$ in an equilibrium system, the probabilities add up:
\begin{equation}
\begin{aligned}
    \tilde{p}_{m'} &=p_m + p_n \\
    &= \dfrac{1}{Z} \exp \left( \ln \left(e^{-\beta E_m} + e^{-\beta E_n}  \right) \right) \\
    &= \dfrac{1}{Z} e^{-\beta F_{m'}}.
\end{aligned}
\end{equation}
In the last expression, we have written the Boltzmann distribution of the coarse-grained state with a free energy instead of an energy. That this term indeed corresponds to a free energy $F_{m'}=\langle E_{mn}\rangle -T S$, given by  the average energy $\langle E_{mn} \rangle $ of the merged states and an internal entropy of the coarse-grained state can be seen as follows: 
Pulling the energy $E_m$ out of the logarithm, we rewrite the free energy of the coarse-grained state as
\begin{equation}
\begin{aligned}
   F_{m'} 
   &= E_m + k_B T \ln  \dfrac{e^{-\beta E_m}}{e^{-\beta E_m}+e^{-\beta E_n}}  \\
   &= E_m + k_B T \ln  \dfrac{p_m}{p_m+p_n}.
   \label{eq:E_m}
\end{aligned}
\end{equation}
Likewise, pulling out $E_n$, we get
\begin{equation}
\begin{aligned}
   F_{m'} &= E_n + k_B T \ln \left( \dfrac{p_n}{p_m+p_n}\right)
   \label{eq:E_n}.
\end{aligned}
\end{equation}
We can define $\Bar{p}_m=p_m/(p_m+p_n)$ and $\Bar{p}_n=p_n/(p_m+p_n)$ as re-scaled probabilities of the microscopic states within the coarse-grained state.
Adding both equations with weights $\Bar{p}_m$ and $\Bar{p}_n$, respectively, yields
\begin{equation}
    \begin{aligned}
         F_{m'} =&  E_m\Bar{p}_m+E_n \Bar{p}_n +k_B T\left( \Bar{p}_m \ln \left( \Bar{p}_m \right) +\Bar{p}_n \ln \left(\Bar{p}_n\right) \right) \\
         =& \langle E_{mn} \rangle - T S,
    \end{aligned}
\end{equation}
where we expressed the free energy in terms of the average energy $\langle E_{mn}\rangle =  E_m\Bar{p}_m+E_n \Bar{p}_n$ of the coarse-grained states and the internal entropy due to lumping two states together which is given by
\begin{align}
    S=- k_B \left[ \Bar{p}_m \ln \left( \Bar{p}_m \right) +\Bar{p}_n \ln \left( \Bar{p}_n \right) \right].
\end{align}
The latter arises due to the uncertainty, which micro-state in the merged state is occupied given that the system is in the coarse-grained state.
The identification of the free energy of the merged states is consistent with the redefinition of the transition rates proposed in equation \eqref{eq:general_transition-rates}. The outgoing transition rates $\tilde{\alpha}_{mj}$ of a merged state $m$ are weighted with with $\Bar{p}_m$ and $\Bar{p}_n$.

\subsection{Comparison with other coarse-graining approaches \label{sec:comparison}}

The coarse-graining approach we describe here is closely related to the so-called local equilibrium approximation \cite{Hummer}. Specifically, the expression we obtain for the rates in the coarse-grained system are equivalent to those obtained based on local equilibrium. However, there are differences in the justification and as a consequence in the applicability of the approach: 
The local equilibrium approximation makes use of time scale separation between degrees of freedom, see e.g.\ \cite{Hummer,Local_EQ_Vilar11081}. A system in a non-equilibrium state often can be described as consisting of  subsystems that are internally in equilibrium. The time evolution of all microstate within a macrostate is assumed to be the rapid \cite{Hummer}. Correspondingly, the local equilibrium approximation requires a time scale separation between rapid transitions between states that will be lumped together and slower transitions to the rest of the system. Typically, the states that are merged will be characterized by approximately equal energy levels.

By contrast, our approach is not based on a time scale separation like the local equilibrium assumption. Instead, we imposed the requirements stated in equations \eqref{eq:probailities_cg} and \eqref{eq:general_transition-rates} to preserve the net numbers of transitions between any pair of coarse-grained states.
Our method thus retains the steady-state probability distribution during coarse-graining. It does not refer to equilibration or to the time scales associated with the dynamics. It can be applied generally to merge arbitrary neighboring states (including those that violate the requirements of the local equilibrium approximation) and will lead to a coarse-grained model that preserves the steady state distribution and fluxes. 
In principle, states with vastly different energy levels and states with a large flux between them can be merged as well. Likewise, transitions between the merged states do neither need to equilibrate rapidly to an equilibrium approximated by a Boltzmann distribution, nor do they need to relax rapidly into a non-equilibrium steady state. However, while our coarse-graining approach preserves the steady state of the system, it necessarily gives an approximation of the dynamics.  The error of this approximation will be smallest if there is a time scale separation, i.e.\ if the requirements of the local equilibrium approximation are fulfilled. In that sense the requirements to preserve steady state quantities are similar to a local equilibrium approximation. However, they still allow the application of the approach in more general cases.  

Several other coarse-graining approaches have been proposed recently \cite{Hummer,SlowFastStates,Vollmer-CG,Altaner_Cycles} that differ in various aspects from each other and from the approach used here. 
In the following, we will briefly compare these approaches with our coarse-graining procedure. Table \ref{tab:Comparison} summarizes this comparison.

In \cite{SlowFastStates},  Pigolotti and Vulpiani presented an adiabatic approximation which eliminates rapidly evolving states from the system. All states with a mean dwell time [the inverse of the exit rate as given by equation \eqref{eq:exit_rates} below] below a certain threshold are removed and the rates of the remaining states are renormalized. The approximation is well suited for system with a time scale separation.
In contrast to our approach, the calculation of a steady-state distribution is not needed. Hence, this approach is suitable e.g.\ for chemical networks with infinite state space, where calculating the steady state distribution is computationally challenging. However, the steady-state distribution is also not preserved in the coarse graining, as the probabilities of the states eliminated by coarse graining are redistributed globally and not only to the neighbouring states.

The approach by Altaner and Vollmer \cite{Vollmer-CG} is based on eliminating 'bridge states', states that can be eliminated in such a way   that the cycle topology and the cycle affinities are preserved. In the coarse-graining step, the steady-state probability of the bridge state is distributed (not necessarily equally) to its two neighbouring states, such that the probabilities of non-neighbour states are unaffected (locality) and the systems' change in system entropy is preserved. 
This method is only applicable if the net flux along the bridge is nonzero. Hence, it is limited to non-equilibrium systems with nonzero currents.
 
In yet another approach \cite{Hummer}, Hummer and Szabo proposed a coarse-graining method that leaves the occupancy-number correlation function $C_{ij}(t)=\langle \theta_i(t)\theta_j(0) \rangle$ unchanged. Here $\theta_i(t)$ is an indicator function that is one if state $i$ is occupied at time $t$ and zero otherwise.
The calculation of the reduced matrix with the transition rates of the coarse-grained system requires matrix inversion and knowledge of the steady-state distribution as in our approach.
However, in general, all transitions rates (not only the outgoing rates from the merged states) are changed. Hence, the method of Hummer and Szabo incorporates non-local changes of the transition matrix.
The reduced matrix of the is computed by Hummer and Szabo in the Markovian limit ($s \rightarrow 0$ in the Laplace space). They call their approach optimal Markov model and compare their model with the local equilibrium approximation which corresponds to $s\rightarrow \infty$ in the Laplace space. If only two states are merged and hence all other macro states contain only one micro state, the local equilibrium approximation is equivalent to the definition of transition rates in equation \eqref{eq:general_transition-rates}.

All coarse-graining procedures discussed so far are based on merging states. A different approach was proposed by Altaner et al.\ in \cite{Altaner_Cycles}. Starting from the fact that edge fluxes can be represented as superpositions of cycle fluxes (see equation \eqref{eq:fluxes_Kinesin} for an example), a mapping is defined that transforms the original graph into a new one. The states in the new graph represent cycles of the original one. The transitions between the states representing cycles are defined in such way that the coarse-grained system fulfills detailed balance.

\begin{table}
\centering%
\caption{Comparison of coarse-graining approaches.\label{tab:Comparison}}
\small
\footnotesize
\scriptsize
\begin{tabularx}{\linewidth}{|X|X|X|X|X|}
\toprule
                                  & Adiabatic approximation with fast and slow states  \cite{SlowFastStates}& Fluctuation preserving coarse-grain method by Altaner and Vollmer in \cite{Vollmer-CG} &  Approach by Hummer and Szabo in \cite{Hummer}& Our transition fluxes preserving coarse-grain approach \\
\midrule
\midrule
Assumption  / required quantities                      & Outside stead- state regime valid too; calculation of steady state not needed                                                                                         &    Steady state; requires non-zero flux along bridge that will be coarse-grained      & Steady state &      Steady state                \\ \midrule
Reduction of state space via        & Bridge                                                                                         &    Bridge  & Lumping together $N$ states in $M$ lumped states                                      &     Merge two nodes                 \\  \midrule
Redefinition of rates               &  Non-local changes of transition rates; affected by the dwell time in the bridge state  &   Proportion of one-way fluxes and proportion of probabilities define new rates from and to the node left and right of the bridge state & Non-local changes of transition rates; Matrix inversion of matrix product such that number correlation function is retained &  Local changes: Outgoing flux from merged nodes preserved; all other rates retained             \\  \midrule
Cycle topology                      & Coarse graining of fundamental cycles possible, no restrictions                                                                                         & Preserved (requirement) & No restrictions                                            &     Coarse graining of branches and fundamental cycles possible                                                                                                         \\  \midrule
Local redistribution of probability &Non local redistribution                                                                                          &  Only the steady-state prob. of the neighboring nodes changed & Probabilities within lumped states add up                                          &  Coarse-grained probability of merged node is the sum; all other prob. preserved                   \\  \midrule
Affinities                & Preserved                                                                                         & Preserved & Not preserved even if topology unchanged                                          &   Changed by the log of proportion of the one-way fluxes between merged states - see eq. \eqref{eq:Kinesin_Affinity}\\  \midrule
Fluxes                     &Not preserved                                                                                         &Preserved &Not preserved                                            &       Preserved              \\ \midrule
Iterative & Commutes& Does not commute& Lumping order not important & Order of coarse-grain iterations not important (commutes) \\ \midrule \midrule
Advantages                          & Well suited for infinite large networks (chemical networks)                                                                                          &  Affinities preserved; can only coarse grain bridges with a non-zero, positive net-flux & Occupancy number correlation function retained                                          &    Coarse-graining of branches, direct mapping between coarse-grained and original states, merging two nodes is iterable and order is not important \\

\bottomrule
\end{tabularx}
\end{table}

Our coarse-graining approach shares some features with these methods, but differs from them in others. For example, our approach requires the calculation of the steady-state distribution of the system, which can be computationally costly for large systems and imposes a limitation on the use of the method. The approach described by Pigolotti and Vulpiani \cite{SlowFastStates} is the only one of the methods presented here that does not require the steady-state distribution.
In our approach, the state space is reduced via the merging of two arbitrary adjacent nodes. Adjacency is the only prerequisite, whereas the method of Altaner and Vollmer \cite{Vollmer-CG} is only applicable for bridge-states such that the cycle topology is preserved.  
The probability of the coarse-grained states is redistributed locally in our approach as well as in  \cite{Vollmer-CG} whereas the methods of \cite{SlowFastStates,Hummer} redistribute the probabilities non-locally.
Likewise, the steady-state fluxes between states are not preserved in the adiabatic approximation with slow and fast states  \cite{SlowFastStates} but are retained in Altaner and Vollmer's \cite{Vollmer-CG} and in our approach.
The cycle affinities (under the condition of preserved network topology) are preserved by the methods in \cite{SlowFastStates,Vollmer-CG}, but not by our approach.
Due to the direct mapping in our coarse-graining approach, one can compare the original and the coarse-grained steady-state distribution. Furthermore, each node retains its meaning due to the direct mapping. A coarse-grained node contains at least two original nodes. 

Table \ref{tab:Comparison} summarizes and compares the coarse-graining approaches with respect to the properties discussed in this section. The comparison indicates that
all coarse-graining approaches have their advantages and disadvantages and are better suited for some applications than others. Our approach is well suited for coarse-graining by removing branches (edges or subgraphs that are not part of any cycle and hence have zero steady-state transition fluxes) and for systems that can make use of the direct mapping between coarse-grained and original states.
Besides, our approach is broadly applicable as it imposes no constraints on the network topology. Furthermore, due to the iterative nature of the approach, we can use it to obtain a hierarchy of models that are coarse-grained to different degrees, a feature that we will use below.

Finally, we want to mention that some authors make a distinction between coarse-graining and model reduction, specifically for systems with continuous variables \cite{Gorban_2006}. The former is accompanied by entropy increase, while the latter is not, e.g. because it reduces a system's dynamics to the slow dynamics on an invariant manifold. For the latter case, related methods for simplifying reaction networks (by eliminating edges based on relations between rates) have been proposed and applied to biochemical systems \cite{Radulescu_2012}.

\subsection{Minimizing the Kullback-Leibler divergence for trajectories\label{sec:KL}}

Another way to motivate the choice of transition rates in the coarse-grained system as given by  equation \eqref{eq:general_transition-rates} is by minimizing the Kullback-Leibler divergence $\text{KL}(q||\tilde{q})$ between probabilities $q$ and $\tilde q$ of trajectories in the original system and in the coarse grained system, respectively  \cite{maxEntropy_Presse}. The Kullback-Leibler divergence \cite{kullback1951} is defined as 
\begin{equation}
\begin{aligned}
\text{KL}(q||\tilde{q})=\Big\langle \ln \left(\dfrac{q}{\tilde{q}}\right)\Big\rangle _q,
\label{eq:KL_div}
\end{aligned}
\end{equation}
and provides a non-symmetric measure of the similarity of two probability distributions.

For this derivation, we write the master equation in matrix notation
\begin{equation}
\begin{aligned}
\dfrac{d \textbf{p}}{dt}&=\textbf{W}\textbf{p},
\end{aligned}
\end{equation}
where the matrix elements of $\textbf{W}$ are given by 
\begin{align}
W_{ij}= \begin{cases}
\alpha_{ji} &\text{ for } i\neq j \\
-\sum_{k\neq i} \alpha_{ik} &\text{ for }i=j.
\end{cases}
\label{eq:exit_rates} 
\end{align}
The diagonal elements are called exit rates and ensure conservation of probability. The exit rate of state $i$ is the sum of all outgoing transition rates. 
The master equation is solved by
\begin{align}
\textbf{p}(t)=e^{\textbf{W}t}\textbf{p}(0),
\end{align}
where $\textbf{G}(t) = e^{\textbf{W}t}$ denotes the propagator. To derive probabilities for trajectories, we discretize time and expanded the propagator for small time steps $\Delta t$: $\textbf{G}=\textbf{1}+\Delta t \textbf{W} $ with
\begin{equation}
\begin{aligned}
G_{ij}&=\Delta t \cdot  W_{ij} \qquad (i\neq j)\\
G_{ii}&= 1-\Delta t \sum_{j\neq i}W_{ji}.\label{eq:G_ij}
\end{aligned}
\end{equation}
The continuous time dynamics described by the original master equation can be recovered in the limit of $\Delta t \rightarrow 0$ \cite{maxEntropy_Presse,Annibale}.
A trajectory $\{i\}=(i_0, i_{\Delta t}, i_{2\Delta t}, ...,i_{t-\Delta t})$ is a sequence of states at different time steps. Its probability can be expressed in terms of the propagator as 
\begin{align}
q(\{i\})=p_{i_0}(0) \prod_{\tau=0}^{t-\Delta t} G_{i_{\tau+\Delta t}, i_{\tau}}.
\end{align}

We now reduce the dimensionality of the state space by merging adjacent states.
Instead of considering a single coarse-graining iteration as in section \ref{sec:CG_procedure}, we can examine a system after several coarse-graining iterations, as the order of those steps does not affect the result. 
After mapping states in the original system (denoted by small letters) to states in the coarse-grained system (denoted by capital letters), we have to define transition rates for the coarse-grained model $\tilde W_{IJ}$. We will show that the choice of transition rates in our coarse-graining approach minimizes the Kullback-Leibler divergence $\text{KL}(q(\{i\})||\tilde{q}(\{ I\}))$, which compares the probability for trajectories in the original and the reduced state space.
A trajectory $\{I\}=(I_0, I_{\Delta t}, I_{2\Delta t}, ...,I_{t-\Delta t})$ in the coarse-grained state space has probability
\begin{align}
\tilde{q}(\{I\})=\tilde{p}_{I_0}(0) \prod_{\tau=0}^{t-\Delta t} \tilde{G}_{I_{\tau+\Delta t}, I_{\tau}},
\end{align}
where the short-time propagator for the reduced system has the matrix elements 
\begin{equation}
\begin{aligned}
\tilde{G}_{IJ}&=\Delta t\, \tilde{W}_{IJ}\qquad (I\neq J)\\
\tilde{G}_{II}&= 1-\Delta t \sum_{J\neq I}\tilde{W}_{JI}
\end{aligned}
\end{equation}
in analogy to equation \eqref{eq:G_ij}.

In order to minimize the Kullback-Leibler divergence in equation \eqref{eq:KL_div}, we have to consider the mean of the logarithm of the distribution $\tilde{q}$ in the coarse-grained system with respect to the distribution $q$ in the original system:
\begin{align}
\langle \ln \tilde{q} \rangle_q=\langle \ln \tilde{p}_{I_0}(0)\rangle_q+  \sum_{\tau=0}^{t-\Delta t} \langle \ln (\tilde{G}_{I_{\tau+\Delta t}, I_{\tau}} )\rangle_q.
\label{eq:Zwischenschritt_ln_p}
\end{align}
If we assume that the distribution $p$ in the original system starts in its steady state, each time interval $(\tau, \tau + \Delta t)$ has the same average, such that
\begin{equation}
\begin{aligned}
\langle \ln \tilde{q} \rangle_q=& \sum_{i_0}p_{i_0} \ln \tilde{p}_{I_0}(0) +\dfrac{t}{\Delta t} \sum_{IJ} \sum_{i \in I, j \in J} G_{ij}p_j \ln (\tilde{G}_{IJ} )
\label{eq:average_p^cg}
\end{aligned}
\end{equation}
the sum over the time steps in equation \eqref{eq:Zwischenschritt_ln_p} can be simplified because $ \langle \ln (\tilde{G}_{I_{\tau+\Delta t}, I_{\tau}} )\rangle_q$ becomes time independent in the steady-state and hence each summand in the sum over time contributes equally. 
The probabilities in a cluster $I$ add up ($p_I=\sum_{i \in I}p_i$) and the cluster propagator can be defined as $G_{IJ}=p_J^{-1} \sum_{i \in I, j \in J}G_{ij}p_j$ such that equation \eqref{eq:average_p^cg} can be written as
\begin{equation}
\begin{aligned}
\langle \ln \tilde{q} \rangle_q=& \sum_{I_0}\tilde{p}_{I_0} \ln \tilde{p}_{I_0}(0)  +\dfrac{t}{\Delta t} \sum_{IJ}  G_{IJ}p_J \ln (\tilde{G}_{IJ} ).
\end{aligned}
\end{equation}
In the limit of long trajectories $t \rightarrow \infty$, the first term can be discarded. In the second term, we separate diagonal ($I=J$) and off-diagonal ($I\neq J$) terms:
\begin{equation}
\begin{aligned}
\dfrac{1}{t}\langle \ln \tilde{q} \rangle_q = & \sum_I G_{II}p_I \dfrac{\ln(\tilde{G}_{II})}{\Delta t}  +\sum_{I\neq J}  \dfrac{G_{IJ}}{\Delta t}p_J \left( \ln(\Delta t) + \ln \dfrac{\tilde{G}_{IJ}}{\Delta t} \right).
\label{eq:before_limit}
\end{aligned}
\end{equation}
Going back to continuous time, i.e.\ for $\Delta t \rightarrow 0$, we can rewrite the factors in equation \eqref{eq:before_limit} such that  
\begin{equation}
\begin{aligned}
\lim_{\Delta t \rightarrow 0} G_{II}&=1 \\
\lim_{\Delta t \rightarrow 0} \dfrac{G_{IJ}}{\Delta t}&= W_{IJ}
\end{aligned}
\end{equation}
and 
\begin{equation}
\begin{aligned}
\lim_{\Delta t \rightarrow 0} \dfrac{\tilde{G}_{IJ}}{\Delta t}&= \tilde{W}_{IJ}\\
\lim_{\Delta t \rightarrow 0} \dfrac{\ln(\tilde{G}_{II})}{\Delta t} &=\lim_{\Delta t \rightarrow 0} \dfrac{\ln(1-\Delta t \sum_{J\neq I}\tilde{W}_{JI})}{\Delta t}
= - \sum_{J\neq I}\tilde{W}_{JI}
\end{aligned}
\end{equation}
where $W_{IJ}$ defines the clustered rates of the original model and $\tilde{W}_{IJ}$ the approximating rates in the coarse-grained model. The term with $\ln(\Delta t)$ cancels with the same term in $\langle \ln q \rangle_q$ in the Kullback-Leibler divergence $\text{KL}(q||\tilde{q})=\langle \ln q \rangle_q- \langle \ln \tilde{q} \rangle_q $.
After performing the limit, equation \eqref{eq:before_limit} can be written as
\begin{equation}
\begin{aligned}
\lim_{\Delta t \rightarrow 0} \dfrac{1}{t}\langle \ln \tilde{q} \rangle_q &= -\sum_I p_I \sum_{J\neq I}\tilde{W}_{JI}+\sum_{I\neq J}W_{IJ}p_J\ln(\tilde{W}_{IJ}) + c \\
&= \sum_{I\neq J} (W_{IJ}p_J \ln(\tilde{W}_{IJ})-\tilde{W}_{IJ}p_J) + c
\end{aligned}
\end{equation}
with a constant $c$ that is independent of of $\tilde{W}_{IJ}$.
The minimization of $\text{KL}(q||\tilde{q})$ is equivalent to maximizing its second term $\langle \ln \tilde{q} \rangle_q $. The derivative with respect to $\tilde{W}_{IJ}$ of the second term $\langle \ln \tilde{q} \rangle_q $ of the Kullback-Leibler divergence is
\begin{align}
\dfrac{d}{d \tilde{W}_{IJ}} \dfrac{1}{t} \langle \ln \tilde{q} \rangle_q = p_J\left( \dfrac{W_{IJ}}{\tilde{W}_{IJ}}-1\right)=0.
\end{align}
Hence, $ \tilde{W}_{IJ} = W_{IJ}$ minimizes the Kullback-Leibler divergence $\text{KL}(q||\tilde{q})$ and yields the same choice for transition rates for the coarse-grained system
\begin{align}
\tilde{W}_{IJ}=W_{IJ}=\dfrac{\sum_{i \in I, j \in J}G_{ij}p_j}{\sum_{j\in J} p_j}
\label{es:Choice_of_rates_KL}
\end{align}
as in equation \eqref{eq:general_transition-rates}.
Summarizing, then, we have shown in this section that minimizing the Kullback-Leibler divergence for trajectories yields the same choice of transition rates as proposed in equation \eqref{eq:general_transition-rates}. 

\section{Coarse graining with preserved cycle topology}
\label{sec:Kinesin}
\subsection{Cycle decomposition of the kinetic diagram for the kinesin motor}
\begin{figure}
    \centering
    \includegraphics{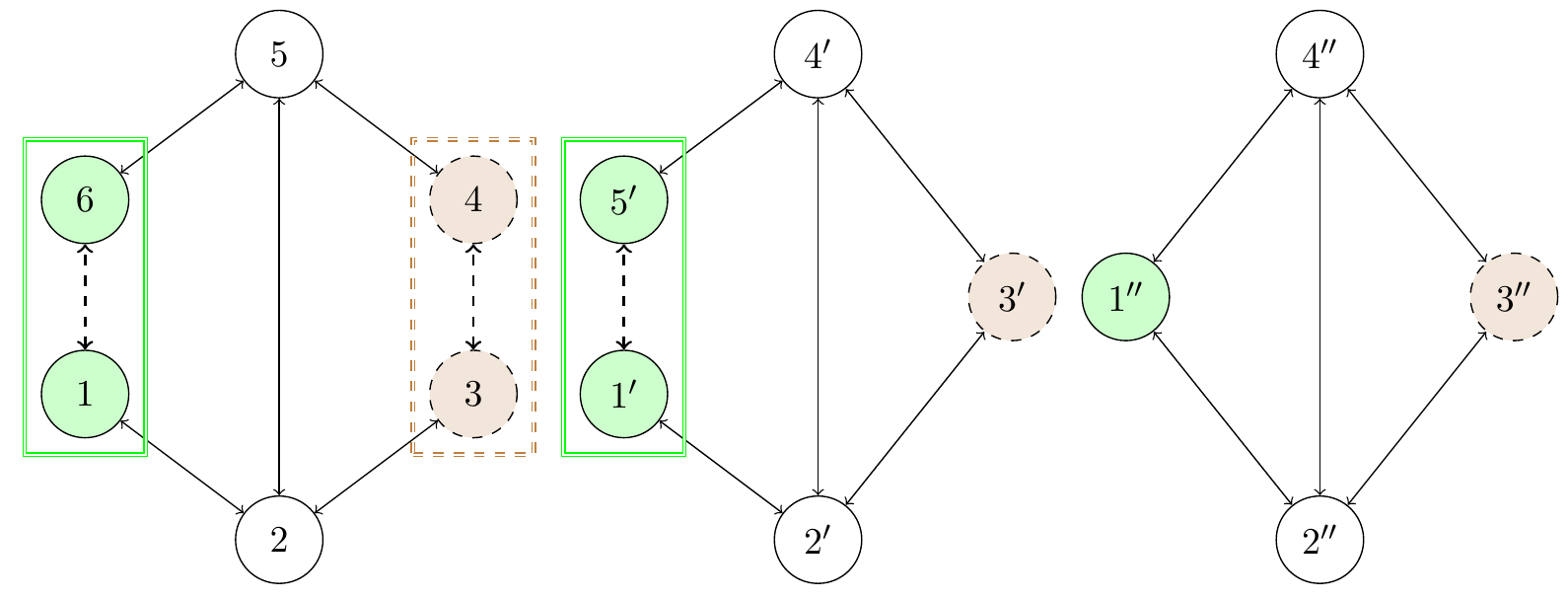}
    \caption{Kinetic diagram for Kinesin from \cite{Liepelt}. The 6-state model will be simplified to a 4-state model without changing the number of cycles within the network. In the forward cycle, states 1 and 6 will be merged to state 1'', while in the backward cycle, states 3 and 4 are described by state 3'' in the coarse-grained network. Except for the fluxes between the merged states, all net transition fluxes are retained. We eliminated the transition with minimal contribution to entropy production in each fundamental cycle. \label{fig:Kinesin-network_CG}}
\end{figure}
In this and the next section, we will discuss the coarse graining of the kinetic diagram of the molecular motor kinesin as an example for simplifying a system in a non-equilibrium steady-state. We will first coarse grain the network without changing its cycle topology and postpone coarse-graining steps to remove cycles to the next section. 

Kinesin is a motor protein with two heads which can carry cargo, moves along microtubule filaments and is powered by the hydrolysis of adenosine triphosphate (ATP) \cite{Woehlke2000, Kinesin_review}.
Kinesin can be described by the 6-states network \cite{Liepelt} that we have already used as an example above, depicted in figure \ref{fig:Kinesin-network_CG}. The six states correspond to different chemical states of the two motor heads (ATP- or ADP-bound or free).  Transition $2\rightarrow 5$ corresponds to a mechanical forward step, the backward transition $5\rightarrow 2$ to a backward step. Alternative mechanical stepping transitions have been discussed \cite{HYEON_KLUMPP}, but will not be considered here. The other transitions describe chemical changes in the motor heads, i.e.\ binding or release of ATP, or ADP and the hydrolysis of ATP to ADP in the front head and rear head (transitions  
$6\rightarrow 1$ and $3\rightarrow 4$, respectively) including the release of phosphate. 

The network can be decomposed into cycles as described in the previous section. The blue-coloured cycle in figure \ref{fig:Cycle_Decomposition} ($C_F=\{1,2,5,6,1 \}$, defined by the dotted chord $1\rightarrow2$)  includes a mechanical forward transition and an ATP hydrolysis transition and is hence called forward cycle $C_F$. It corresponds to the normal working cycle of the kinesin motor under low loads. 
The red coloured cycle $C_B=\{2,3,4,5,2 \}$ (defined by the dashed chord $2\rightarrow 3$) incorporates a mechanical backward transition and an ATP hydrolysis transition and is called backward cycle $C_B$. The third cycle, which is not fundamental in the decomposition in figure \ref{fig:Cycle_Decomposition}, contains no mechanical transitions but two ATP hydrolysis transitions and hence is a dissipative or futile cycle $C_D$.
The network has two independent transition fluxes. With the choice of two chords in the cycle decomposition depicted in figure \ref{fig:Cycle_Decomposition}, all transition fluxes can be expressed in terms of the chord fluxes $J_{12}$ and $J_{23}$ as
\begin{equation}
\begin{aligned}
J_{12} &= J_{56} = J_{61} = J_F+J_D \\
J_{23} &= J_{34} = J_{45} = J_B+J_D \\
J_{25} &= J_{12}-J_{23}=J_F-J_B.
\label{eq:fluxes_Kinesin}
\end{aligned}
\end{equation}
For the sake of completeness, the relations between transition and cycle fluxes are listed as well. The cycle fluxes $J_C$ (with $C=F,B,D$) are defined according to Hill's convention \cite{hill} and can be interpreted as the number of completions of cycle $C$ per time. Cycle fluxes will be needed later to calculate fluctuations of steady-state quantities. Equation \eqref{eq:4nodes3cycles-cylesfluxes} in the supplementary material, shows how to calculate the cycle fluxes in the kinesin network.
The entropy change after completion of a cycle $C$ can be expressed in terms of the transition affinities within the cycle as defined in equation \eqref{eq:Cycle_Affinity}. With the choice of chords depicted in figure \ref{fig:Cycle_Decomposition}, the affinity of the futile cycle $C_D$ can be written as a linear combination of the fundamental cycles as
\begin{align}
\Delta S_D&=\Delta S_F+\Delta S_B .
\end{align}
The entropy production in a system in a non-equilibrium steady-state can be written in terms of fundamental cycle affinities and the flux through the fundamental cycles defining chords as stated in equation \eqref{eq:entropy_prod_decomposition}. This yields
\begin{equation}
\begin{aligned}
P&=\sum_l J_l \Delta S_{C_l}\\
&= J_{12}\Delta S_F+J_{23} \Delta S_B.
\end{aligned}
\end{equation}
\subsection{Coarse graining of the kinetic diagram of kinesin without changing the network topology \label{sec:Kinesin_preserved}}
To simplify the state space, we merge states 3 and 4 in the backward cycle, and subsequently states 1 and 6 in the forward cycle as shown in figure \ref{fig:Kinesin-network_CG}. The network topology, i.e. the number of fundamental cycles, remains unchanged by these network simplifications. The coarse-grained network still contains three cycles, of which two are fundamental.
By construction of our coarse-graining approach, all transition fluxes are conserved. In particular, the chord fluxes remain unchanged. ($\tilde{J}_{1''2''}=J_{12}$ and $\tilde{J}_{2''3''}=J_{23}$). The flux between the merged nodes is lost however and hence the fundamental cycle affinities are reduced by the transition affinity of the eliminated transition between the merged nodes:
\begin{equation}
\begin{aligned}
\Delta \tilde{S}_F &= \Delta S_F - \ln  \dfrac{\alpha_{61}p_6}{\alpha_{16}p_1}  \\
\Delta \tilde{S}_B &= \Delta S_B - \ln  \dfrac{\alpha_{34}p_3}{\alpha_{43}p_4}.
\label{eq:Kinesin_Affinity}
\end{aligned}
\end{equation}
The cycle affinities are retained if the net flux between the merged states is zero.
Hence, the coarse graining also reduces the entropy production by
\begin{equation}
\begin{aligned}
\Delta P&=P-\tilde{P} \\
&= J_{12} \ln  \dfrac{  \alpha_{61}p_6}{\alpha_{16}p_1} +J_{34} \ln  \dfrac{\alpha_{34}p_3}{ \alpha_{43}p_4},  
\label{eq:dif_entropy_prod_Kinesin}
\end{aligned}
\end{equation}
unless detailed balance is fulfilled between the merged states.

\subsection{Fluctuations in the kinesin network}
\begin{figure}
    \centering
    \includegraphics{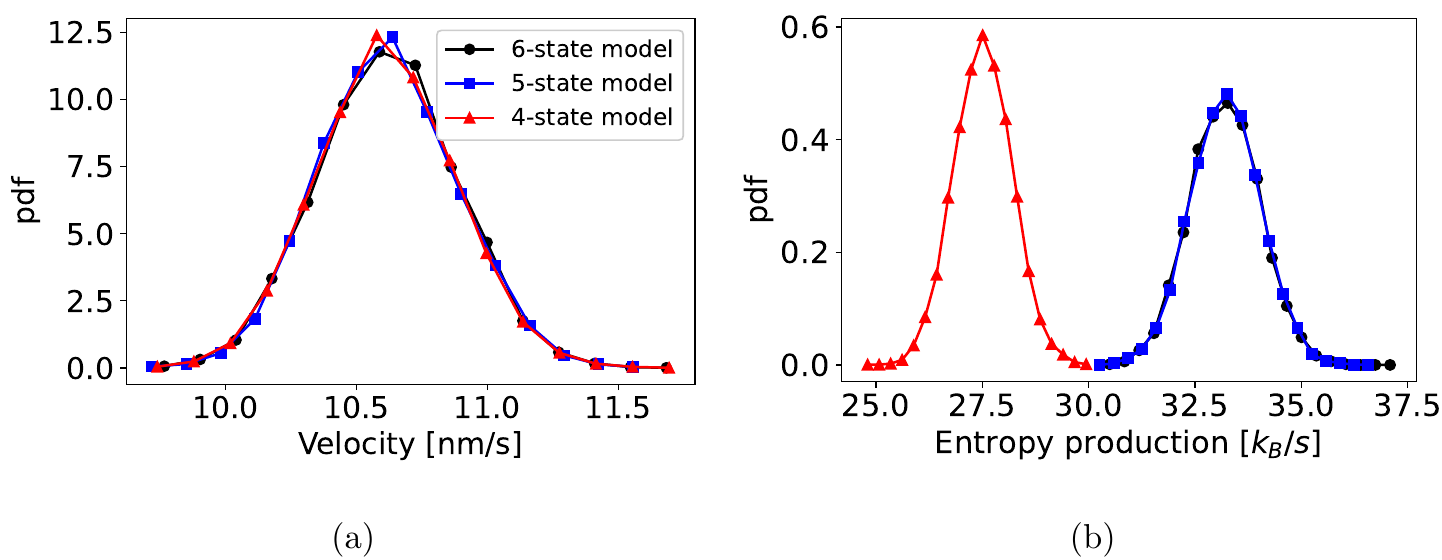}
    \caption{Simulation results for the velocity (a) and the entropy production (b) of a kinesin motor. The rate constants for the 6-state model are from \cite{Liepelt} for chemical concentrations $[ATP]=[ADP]=[P]=1\, \mu M$, stepping size $l=8\,$nm  and an external load force $F=1$\,pN. 10000 trajectories have been sampled for each model with a simulation time $\tau=1200$\,s. The black data (circles) corresponds to the original 6-state model from \cite{Liepelt} which is depicted in figure \ref{fig:Kinesin-network_CG}. The blue data (squares) represents a coarse-grained system with five states, where the original states 3 and 4 have been merged. The four-state model is depicted in red and with triangles. 
The entropy production for a trajectory of finite length was calculated as in equation \eqref{eq:entropy_prod_trajectory}. For the entropy production, the 4-state model underestimates the mean by around 17\%. The velocity $v$ of a kinesin motor is proportional to the step size $l=8$\,nm and to the flux along the mechanical transition ($J_{25}$ in the 6-state, $J_{2'4'}$ in the 6-state and $J_{2''4''}$ in the 4-state model). Our coarse-graining approach preserves the mean of the velocity whereas it does not preserve variances of observables. However, the numerical differences for the variance of the velocity are very small.
}
    \label{fig:PDF_Kinesin}
\end{figure}
Our coarse-graining approach is designed to preserve fluxes, which correspond to averages of observable properties of the kinesin motor. 
To demonstrate how our coarse-graining approach affects the distribution of quantities such as the entropy production and velocity of kinesin, we simulated 10000 trajectories of finite duration (analytical results are presented in appendix \ref{sec:Fluctuations_explictly}).  We compared the original 6-state model of Liepelt and Lipowsky \cite{Liepelt} with two coarse-grained systems. In the 5-state system, state 3 and state 4 of the original model are merged such that the backward cycle loses one transition. In the 4-state model, states 1 and 6  are merged in addition, such that the forward cycle loses one transition as well.  

In figure \ref{fig:PDF_Kinesin}(a), we show the velocity distribution for the 6-, 5-, and 4-state model obtained from 10000 trajectories with simulation time 1200\,s. The velocity of the motor is proportional to the flux between states 2 and 5 and is given by
\begin{align}
    v = l J_{25},
\end{align}
where $l$ is the step size of the motor.
The variance of the velocity decreases for longer trajectories. The velocity is proportional to the net flux of the mechanical transition $2\rightarrow 5$ in the original model and to the respective mechanical transitions in the coarse-grained models ($2'\rightarrow 4'$ in the 5-state model, $2''\rightarrow 4''$ in the 4-state model). Leaving the network topology unchanged, our coarse-graining approach preserves the steady-state flux and hence the mean velocity ($v=\tilde{v}=10.689$\,nm/s) remains unchanged in the course of the coarse-graining procedure. Moreover, the full velocity distribution seems to be unaffected by the coarse graining. But in fact, our coarse-graining approach does not preserve variances of observables. However, by calculating the variance analytically, one can see that the difference and hence the error due to coarse-graining is very small in this case: For the 6-state model, the standard deviation of the velocity is $\sigma_v=0.2669482111$\, nm/s and for the 4-state model $\sigma_{\tilde{v}}=0.2669482115$\, nm/s. Coarse graining from six to four states only leads to a relative increase of the standard deviation of $3 \cdot 10^{-9}$. In general, only mean values are conserved by our reduction scheme. However, in this example the relative differences are small because the network topology is preserved and the flux between the merged nodes is small compared to the flux along the mechanical transition between states 2 and 5.

The corresponding distributions for the entropy production for trajectories of length 1200\,s are shown in figure \ref{fig:PDF_Kinesin}(b). For a trajectory of duration $\tau$ with $m$ transitions $n_0\rightarrow n_1 \rightarrow ... \rightarrow n_{m-1} \rightarrow n_m$ we consider the quantity 
\begin{align}
P_{\tau}=\dfrac{k_B}{\tau}\ln \dfrac{\alpha_{ n_1}\alpha_{ n_2} ...\alpha _{n_{m-1}n_m}}{\alpha _{n_1 n_0}\alpha _{n_2 n_1} ...\alpha _{n_{m}n_{m-1}} }
\label{eq:entropy_prod_trajectory}
\end{align}
as entropy production as proposed by Lebowitz and Spohn in \cite{Lebowitz_1999} and also used in e.g. \cite{Puglisi_2010}.
From equation \eqref{eq:dif_entropy_prod_Kinesin}, we expect the mean of the entropy production for the coarse-grained systems (4-state model in red and 5-state model in blue) to be reduced compared to the 6-state model (black). The variance is not retained either in the coarse-graining step. However, the numerical differences between the 6-state and the 5-state model are very small, as the edge with the minimal contribution to entropy production in the backward cycle is removed and the backward cycle is dominated by the forward cycle for small loads. Coarse graining from six to five states leads to a relative decrease of the mean of the entropy production of about 0.02\%.
Hence the 6-state and the 5-state model show similar distributions.
In general, however, mean and variance of entropy production are not retained by our coarse-graining approach.
Coarse graining the edge with minimal contribution in the forward cycle reduces the mean entropy production by about 17\% compared to the original model. 
That the latter coarse-graining step has a stronger impact is related to the fact that the forward cycle is dominant and yields the largest contribution to entropy production, at least under the conditions of low external force used here. With increasing forces, the backward cycle becomes more important. The force dependency will be discussed in more detail in the next section, when multi-cyclic and uni-cyclic kinesin models will be compared. 

To summarize, the 6-state model for kinesin has been simplified without changing its network topology. The steady-state velocity remains unchanged if the network topology is preserved. Whereas, the mean of the entropy production always decreases if a transition is eliminated that does not fulfil detailed balance.
Variances in general (here for entropy production and velocity) are not preserved. 

\section{Coarse graining with changed cycle topology: Iteratively coarse-grained kinesin\label{sec:Changed_cycle_typology}}
\subsection{General approach}

In the previous section, we coarse grained the kinetic diagram for kinesin without changing the cycle topology: Both the 6-state model of Liepelt and Lipowsky \cite{Liepelt} and the coarse-grained 4-state model contain three cycles. However, the kinesin literature, in particular earlier studies, also contains models with a single cycle. One prominent example is the uni-cyclic 4-state model by Fisher and Kolomeisky \cite{Fisher}.
In this section, we will also allow for the removal of cycles in coarse graining, which allows us to obtain such a uni-cyclic model from the 6-state model with three cycles. 
We use a general approach, in which we iteratively eliminate states from the kinesin model to simplify the description of the molecular motor and to obtain a hierarchy of models with different levels of coarse-graining.
In each coarse-graining iteration, the two states that are merged are chosen such that the transition or the cycle with minimal entropy production is removed and thus 
the difference in entropy production between the models is minimal.
In general, there will be coarse-graining steps that preserve the cycle topology and steps that change it.
For the kinesin model, we can perform four iterations of the coarse-graining step. In each iteration, the model loses one state, such that we end up with a 2-state network after step 4. 
The general coarse-graining algorithm is as follows:
\begin{enumerate}
    \item Find the (chemical) transition $(ij)$ with minimal contribution to the entropy production: 
    \begin{align}
        \min_{(ij)}J_{ij}\Delta S_{ij} = \min_{(ij)} (\alpha_{ij}p_i-\alpha_{ji}p_j) \cdot \ln \dfrac{\alpha_{ij}p_i}{\alpha_{ji}p_j} .
        \end{align}
        The restriction to chemical transitions is not generally necessary, but for the specific example of the kinesin motor, it guarantees that stepping remains part of all coarse-grained models.
    \item After finding the transition $(ij)$ with minimal contribution to entropy production, we have to check whether merging states $i$ and $j$ results in a change in cycle topology. If the system loses a cycle $C$ due to merging states $i$ and $j$, one checks whether the contribution to the entropy production $J_{ij}\Delta S_C$ is smaller than any contribution to the entropy production $J_{mn}\Delta S_{mn}$ for any transition $(mn)$ which is not part of cycle $C$, i.e. one checks if 
    \begin{align}
        J_{ij}\Delta S_C <  \min_{(mn)\not\in C} J_{mn}\Delta S_{mn}.
        \label{eq:check_minimal_contribution}
    \end{align}
    \item If equation \eqref{eq:check_minimal_contribution} is fulfilled or merging states $i$ and $j$ does not change the cycle topology, merge the pair of states $i$ and $j$. If equation \eqref{eq:check_minimal_contribution} is not fulfilled and merging states $m$ and $n$ does not change the cycle topology, merge states $m$ and $n$. Otherwise, eliminate the cycle with minimal contribution to the entropy production.
    \item Repeat until one obtains an equilibrium two-state network.
\end{enumerate}

We will demonstrate how this algorithm can be used to simplify the model for a kinesin motor under different external load forces. For a given value of the force, the coarse-grained models at different levels of the hierarchy will be compared with respect to their network topology and the distributions of their entropy production and velocity. Kinesin preferably moves in the forward direction when no load force is applied. Its velocity decreases under load \cite{Visscher1999,Carter2005}; the force that results in a zero mean velocity is called stall force (approximately 7\,pN \cite{Carter2005}). We consider three cases: (a) no load (section \ref{sec:Iterativ_Force=0}), (b) a load below the stall force (section \ref{sec:Iterativ_Force=5}), and (c)  a load above the stall force, for which the motor moves backwards (section \ref{sec:Iterativ_Force=8}).

Since the coarse-graining procedure we use generates a hierarchy of models, one can ask which degree of coarse graining is optimal in some sense. It is desirable to remove as many states as possible, while retaining as much as possible the key characteristics of the model. We propose to use the entropy production as the characteristic that should be retained as much as possible, as it quantifies the out-of-equilibrium nature of the system. Entropy production is reduced with every step of coarse graining. After four iterations, four states have been eliminated such that kinesin is described by a two-state model, which is in equilibrium by definition and has zero entropy production. To balance the reduction in the number of nodes in the network and the loss of entropy production, we introduce a cost function (a free energy-like quantity)
\begin{align}
F(\nu)=\dfrac{\Delta P(\nu)}{P_0}-T \dfrac{\Delta N(\nu)}{N_0}.
\label{eq:CG_Criterion}
\end{align} 
Here the change in the number of nodes $\Delta N= N_0-N(\nu)$ and the change of entropy production $\Delta
P=P_0-P(\nu)$ are both determined with respect to the original 6-state model and normalized to the respective value in the original model denoted with subscript 0. We have lost a fraction $\Delta P(\nu )/P_0$ of the entropy production and eliminated a fraction $\Delta N(\nu )/N_0$ of the states after coarse-graining iteration $\nu$.
The weighting factor $T$ controls the relative weight of the information loss (the entropy production) and the simplicity (the number of eliminated states) of our model. In the following, we chose $T$ to be one and hence both terms contribute equally. We can then find the optimal level of coarse-graining by minimizing $F(\nu )$. 

\subsection{Different levels of coarse-grained models for zero external load force\label{sec:Iterativ_Force=0}}
\begin{figure}
    \centering
    \includegraphics{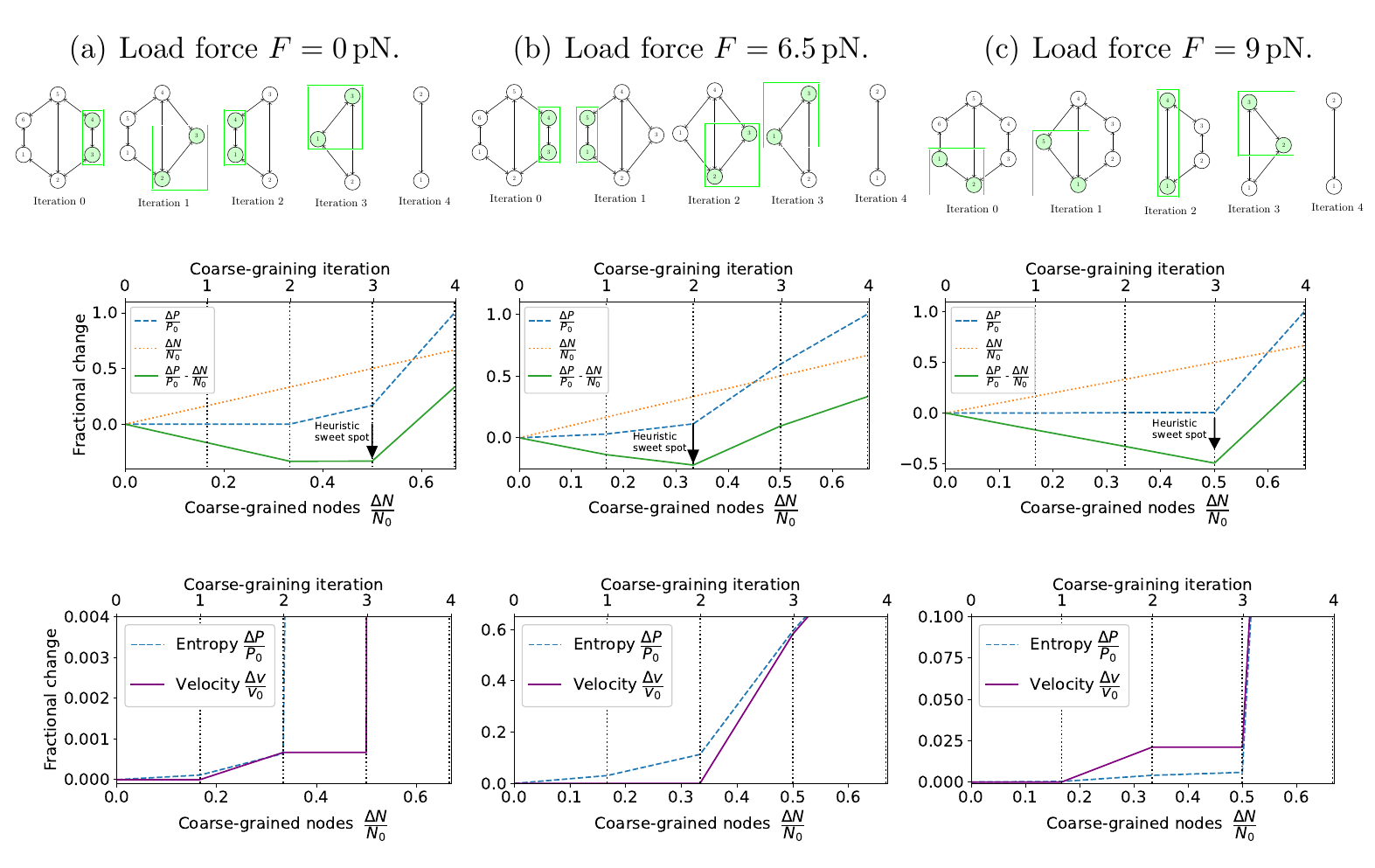}
    \caption{Different models for a kinesin motor. In (a), no external stall force is applied and the forward cycle is dominant. In (b), the kinesin motor is at an external load force of 6.5pN, which is a substantial force, but smaller than the stall force. In (c), the applied load force is larger than the stall force and the backward cycle is dominant.}
    \label{fig:Iterativ_F=0,F=8}
    \label{fig:Iterativ_F=5}
\end{figure}
For a free kinesin motor with zero external force, figure \ref{fig:Iterativ_F=0,F=8}(a) shows the hierarchy of models at different levels of coarse-graining. A uni-cyclic 3-state model (after three iterations) represents a sweet spot between information loss in terms of entropy production and simplicity of the model if both terms are equally weighted ($T=1$). 
If no external force is applied, the forward cycle is dominant and yields the largest contribution to the entropy production $P= J_{12} \Delta S_F+J_{23} \Delta S_B$. Therefore, the first two coarse-graining iterations only affect the backward cycle. The first step preserves the network topology and the backward cycle loses the transition with the smallest contribution $J_{ij}\Delta S_{ij}$ to entropy production. This does not affect the two independent transition fluxes $J_{12}$ and $J_{23}$ (chord fluxes). Hence, the mean velocity of the motor is not changed either. 
The second iteration changes the network topology by eliminating the backward cycle. Thus, the coarse-grained model after two steps is  a uni-cyclic network. The flux in the uni-cyclic network equals the chord flux $J_{12}$ in the 6-state model. Losing one fundamental cycle and one independent transition flux changes the mean velocity of the motor.
Iteration three eliminates a transition within the uni-cyclic 4-state network. The last coarse-graining iteration results in a two-state network which obeys detailed balance by construction. Hence, both the entropy production and the mean velocity of the motor are zero.

In summary, the largest changes in entropy production and mean velocity result from the last coarse-graining step. For zero or small external load forces, the forward cycle is dominant. The backward cycle can be removed by coarse graining, because the approximation error in the steady-state velocity is smaller than 1\%. 

The uni-cyclic 3-state model represents a sweet an optimal trade-off between information loss and simplicity of the model if the motor works at zero external load force.  To investigate whether the coarse-grained model approximates the original model well also in the presence of a load force, we plot the differences in the velocity and the entropy production between various coarse-grained models and the original model as functions of the force in figure \ref{fig:Kinesin_distribution} (the corresponding variances are shown in figure \ref{fig:distributions_variance} in the appendix).

Both figures indicate that uni-cyclic models approximate the original model well for zero or small external load forces. For increasing forces, the approximation of the velocity and the entropy production distribution becomes less accurate. The next subsection shows that changing the force parameter results in a different hierarchy of models at different levels of coarse-graining.

\subsection{Different levels of coarse-grained models for an external load force smaller than the stall force\label{sec:Iterativ_Force=5}}
With an increasing external load force, the mean velocity of the molecular motor decreases and the forward cycle becomes less dominant in terms of entropy production. Figure \ref{fig:Iterativ_F=5}(b) shows different models for a kinesin motor at an external load force of 6.5\,pN, which is a substantial force, but smaller than the stall force. The mechanical transition $2\rightarrow 5$ has the smallest contribution to the steady-state entropy production. We do not merge states 2 and 5 to preserve the biological interpretation of mechanical steps. If we consider only chemical transitions, merging two states in the forward cycle minimizes the change in entropy production after the first coarse-graining step. In the second iteration, the backward cycle loses a transition such that the network topology has not been changed after two steps. Therefore, the system still incorporates two fundamental cycles and two independent transition fluxes and the mean velocity remains unchanged. The 4-state model after two iterations of coarse graining is the same model that was already described in section \ref{sec:Kinesin_preserved} and shown in figure \ref{fig:Kinesin-network_CG}.
The backward cycle is lost after iteration 3. The change of the network topology results in a change of the mean velocity and a kink in the difference of entropy production with respect to the 6-state model. As before, the last iteration yields a 2-state model which obeys detailed balance by definition and has zero mean velocity and entropy production.

If the change in entropy production and the number of coarse-grained nodes are equally weighted, the sweet spot between information loss and simplicity can be found after coarse-graining iteration 2 as depicted in figure \ref{fig:Iterativ_F=5}(b). The resulting tri-cyclic 4-state model preserves the stall force and has an approximation error of 12\% for the steady-state entropy production. The uni-cyclic 3-state network after iteration 3 exhibits pronounced differences in the velocity and entropy production compared to the original 6-state model as depicted in figure \ref{fig:Kinesin_distribution}(b). For zero or small external load forces, the changes in these observables are very small for a uni-cyclic model  as discussed in the previous section. But with increasing forces acting on the motor, the differences become more pronounced for uni-cyclic models and the tri-cyclic 4-state network becomes the better approximation of the original 6-state model, as it  preserves the network topology of the original model.
This is an example for how changes in a parameter that modify the transition matrix can affect the results of our coarse-graining algorithm.

\subsection{Different levels of coarse-grained models for an external load force larger than the stall force\label{sec:Iterativ_Force=8}}
Finally, we consider a force above the stall force. 
In figure \ref{fig:Iterativ_F=0,F=8}(c), the 6-state model for the motor is iteratively coarse grained for an external load force $F=9$\,pN.
In this situation, the molecular motor moves backwards and the mean velocity is negative. The backward cycle is dominant. The transitions in the forward cycle contribute less to the entropy production than those in the backward cycle. Hence in the first coarse-graining step, a transition in the forward cycle is removed and in the second step, the forward cycle is eliminated entirely. The mean velocity is changed in those steps in which the network topology is modified.

\subsection{Summary}
\begin{figure}
    \centering
     \includegraphics{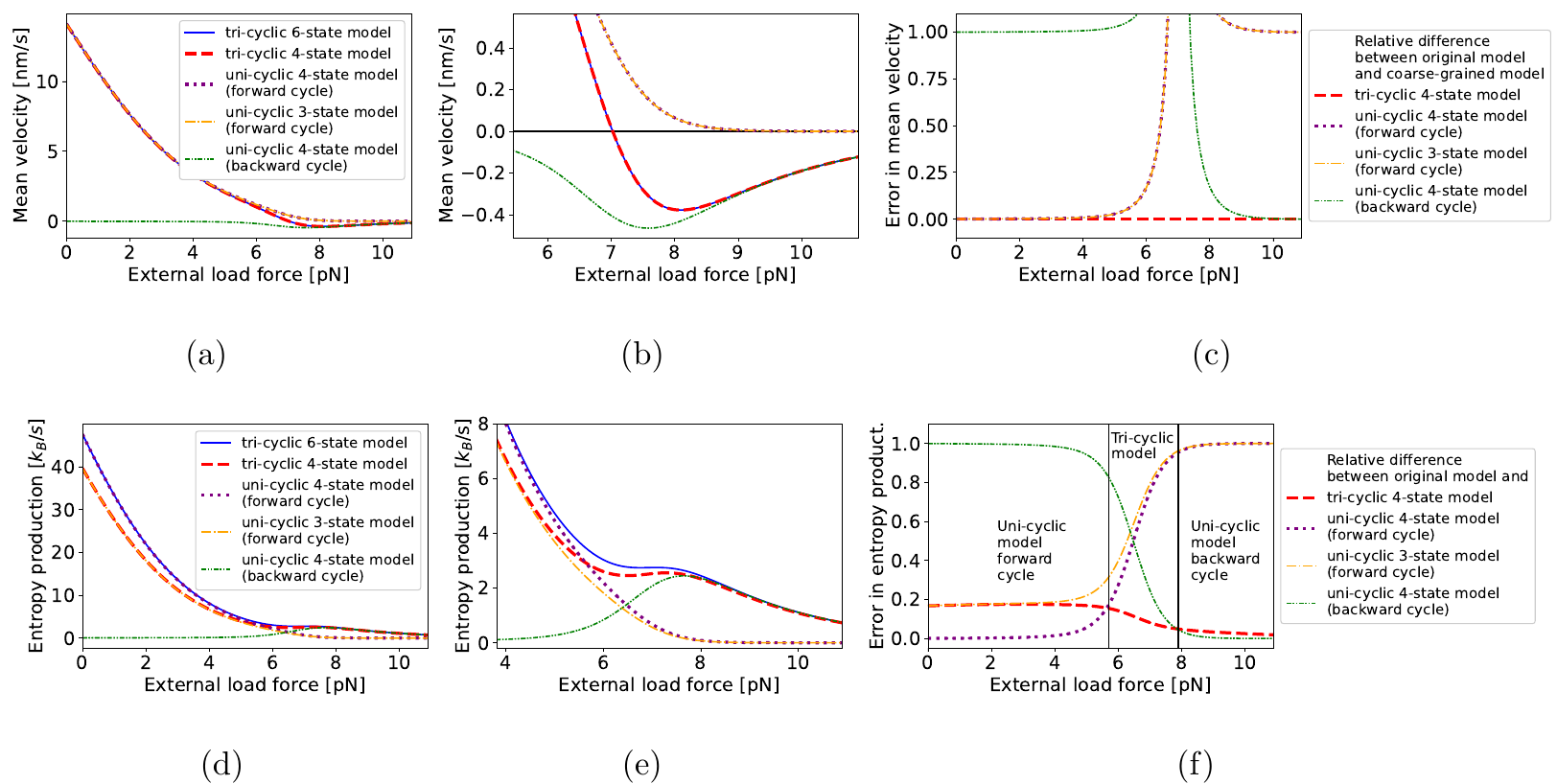}
    \caption{In (a) and (d), the mean velocity and entropy production are plotted against the external load force acting on the motor for different kinesin models, see figure \ref{fig:Iterativ_F=0,F=8}. The figures in the second column, (b) and (e), show an enlarged view of the differences between tri-cyclic and uni-cyclic systems for large external load forces. Furthermore, the relative differences between the mean velocity and the entropy production for coarse-grained models and the original 6-state model respectively are depicted in the third column in (c) and (f). The tri-cyclic 4-state model is optimal in terms of preserving the velocity of the motor. Moreover, the load force ranges indicated in (f) show which model is most suitable for approximating the entropy production within the given range. The motor has zero mean velocity for $F=7$\,pN. For no or small external load forces, uni-cyclic models resembling the forward cycle approximate the mean velocity and entropy production well with small deviations. For load forces around the stall force, on the other hand, a coarse-grained model is better provided it has the same network topology as the original model. In figure \ref{fig:distributions_variance} in the supplementary material, the variances of velocity and entropy production are plotted against the external load force. \label{fig:Kinesin_distribution_entropy}\label{fig:Kinesin_distribution}}
    \label{fig:Kinesin_distribution_velocity}
\end{figure}
In the previous sections, the 6-state model for kinesin with two fundamental cycles has been iteratively coarse-grained such that the difference in the steady-state entropy production is minimal. 
In addition, we determined the velocity as a function of force, the main quantity of interest from an experimental point of view. 
Plotted in figure \ref{fig:Kinesin_distribution}(c) and (f), the relative differences between these quantities in coarse-grained models and  in the original 6-state models show a pronounced difference between models that preserve the network topology and models that do not. As long as the network topology is preserved, the velocity is unchanged by the coarse-graining and the entropy production is only weakly affected. However, when the network topology is modified in a coarse-graining step, both quantities change. Similar observations pertain to the variances of these quantities, as shown in Appendix \ref{sec:Fluctuations_explictly}. 
The difference between coarse-graining with and without preserved network topology is particularly pronounced for forces around the stall force, where multiple cycles contribute to the dynamics. 
Uni-cyclic models are a good approximation for load forces that are considerably smaller (or larger) than the stall force. For small forces, the motor walks in the positive direction and has a dominant forward cycle, such that the backward cycle can be neglected. At and above the stall force, however, the backward cycle makes a strong contribution. 
The uni-cyclic model that resembles the backward cycle approximates the velocity of the motor well if the load force is considerably larger than the stall force and hence the backward cycle is dominant such that the forward cycle can be neglected.
We also note that since changes in network topology affect the velocity, they shift the stall force itself. The new stall force is $\Tilde{F}_{stall}=13.9$\,pN when coarse-graining has removed the backward cycle (a consequence of this is that a different parametrization is  needed if a unicyclic model is used to describe the kinesin motor). Furthermore, the entropy production becomes zero at the stall force in uni-cyclic networks, while it remains finite in the 3-cycle model, reflecting the fact that in the uni-cyclic model the stall condition produces an equilibrium state \cite{HYEON_KLUMPP}.

Different external load forces yield different coarse-graining paths: The forward cycle is dominant for small or no forces. The transitions within the forward cycle have a larger contribution to entropy production than the one in the backward cycle. Therefore, eliminating transitions in the backward cycle results in small changes in entropy production. 
While for small forces the network topology is changed in the second coarse-graining step, it is only affected in the third step close to the stall force. 
For forces above the stall force, the  backward cycle is dominant and thus, the uni-cyclic model obtained from coarse graining is based on the backward rather than the forward cycle. 
In figure \ref{fig:Kinesin_distribution}(c) and (f), we can distinguish between load force ranges. For load forces smaller than 5.7\,pN, the forward cycle is dominant and the approximation error for the entropy production is minimized by a uni-cyclic model which resembles the forward cycle, cf.\ figure \ref{fig:Iterativ_F=0,F=8}(a). For load forces 5.7\,pN $< F <$ 7.9\,pN, the tri-cyclic 4-state model depicted in figures  \ref{fig:Kinesin-network_CG} and \ref{fig:Iterativ_F=5}(b) minimizes the difference in entropy production between the original 6-state model and a coarse-grained model.
For forces $F>7.9$\,pN, the backward cycle is dominant and the uni-cyclic model resembling the backward cycle from figure \ref{fig:Iterativ_F=0,F=8}(c) yields the smallest difference in entropy production with respect to the original model. However, uni-cyclic models do no preserve the stall force in the system and yield higher approximation errors for the velocity of the motor. Figure \ref{fig:Kinesin_distribution}(c) indicates that the tri-cyclic 4-state model is optimal in terms of preserving the velocity of the motor.

Generalizing from the example of kinesin, we can state the observation that changing the network topology yields larger changes in the distribution of observables than coarse-graining that preserves the network topology, if no cycle is dominant. As seen before, the mean and variance of observables can be expressed in terms of fundamental-cycle fluxes. Changing the cycle topology goes along with a change in the number of fundamental cycles which can be seen as a basis in which the behaviour of the system can be expanded.
To give an illustrative example, we once again consider the entropy production, which can be calculated in terms of fluxes and affinities of all transitions (ij) in a system:
\begin{align}
    P=\sum_{(ij)}J_{ij}\Delta S_{ij}.
\end{align}
If the system contains $N$ states and $E$ transitions, the sum has $E$ terms.
Equivalently, the entropy production can be also expressed as a sum of fewer terms, namely $L=E-N+1$ (equation \ref{eq:Number_chords}), if we make use of the cycle decomposition. 
Reducing the number of fundamental cycles $L$ thus always reduces the number of terms $L$ in the expansion of the entropy production in the fundamental cycles basis. If on the other hand the network topology is preserved, the chord fluxes $J_l$ are unchanged whereas the cycle affinities $\Delta S_{C_l}$ of the fundamental cycles $C_l$ are changed and the number of terms is preserved. 

\section{Discussion and Conclusion}
We have introduced a coarse-graining procedure for systems governed by a master equation. The approach merges two states such that the probability of the coarse-grained state is the sum of probability of the two fine-grained states by redefining the transition rates according to equation \eqref{eq:general_transition-rates}. All probabilities of states not affected by the merge are retained. The outgoing transition rates of the coarse-grained state are chosen such that the outgoing flux is conserved.
This approach changes steady-state probabilities and transition rates only locally.
The procedure can be applied iteratively and has no constraints on the network topology. This is in contrast to the coarse-graining approach presented by Altaner and Vollmer \cite{Vollmer-CG}, which is restricted to merging adjacent bridge states, hence coarse-graining of branches or states with more complicated transitions to the rest of the network - as the one depicted in figure \ref{fig:Bsp_Coarse_graining} - is not possible. The only constraint for our approach is the adjacency of states that shall be merged. 
As a consequence, the presented approach can (and typically will) reduce the number of fundamental cycles in the system as shown in section \ref{sec:Changed_cycle_typology}.
 
Moreover, we propose that for an iterative coarse-graining procedure, balancing the (unwanted) loss of entropy production with the (wanted) reduction of the number nodes of the network can result in a heuristic sweet spot of an optimally coarse-grained model.
The cost function proposed in equation \eqref{eq:CG_Criterion} quantifies the balance of information loss and simplicity of the model. 
The relative weight parameter $T$ is so far chosen arbitrarily. With $T=1$, the information loss (the entropy production) and the simplicity (the number of eliminated states) contribute equally.
In the limit of $T\rightarrow 0$, no coarse graining at all is favourable and the minimum of the cost function is always the original model. Whereas in the limit of $T\rightarrow \infty$, the minimum of the cost function corresponds to a 2-state equilibrium model where the maximal number of states have been merged.
While the iterative coarse-graining procedure is not restricted by the features of the network, the criterion for optimal coarse-graining can only be applied to systems out of thermodynamic equilibrium due to the use of the loss in  entropy production as a measure of information. Systems in equilibrium have zero entropy production and hence the criterion cannot be used to compare models that are coarse-grained to different levels for equilibrium systems. 

In section \ref{sec:KL}, we minimized the Kullback-Leibler divergence for trajectories to justify our choice of transition rates in the coarse-grained system.
The Kullback-Leibler divergence is a special case of the more general $\alpha$-divergence 
\begin{equation} \begin{aligned} D_{\alpha}(q||\tilde{q}) &=\dfrac{1}{\alpha(1-\alpha)}\left\langle 1- \left( \dfrac{q}{\tilde{q}} \right) ^{\alpha} \right\rangle _{\tilde{q}}.
\label{eq:alpha-divergence}\end{aligned}\end{equation}
The Kullback-Leibler divergence is not symmetric: $KL(q||\tilde{q}) \neq KL(\tilde{q}||q)$, whereas the $\alpha$-divergence is symmetric for $\alpha=1/2$. Minimizing the $\alpha$-divergence for $\alpha=1/2$ or an arbitrary $\alpha$ could lead to interesting other definitions of transition rates in the coarse-grained system. 

\appendix

\section*{Appendix }
\subsection{Parametrisation for the molecular motor kinesin\label{sec:Kinesin_Parametr_Liepelt}}
Liepelt and Lipowsky proposed a 6-state model for Kinesin with transition rates
\begin{align}
\alpha_{ij}=\alpha_{ij}^0 I_{ij}([X])\Phi_{ij}(F),
\label{eq:parameter_kinesin}
\end{align}
where the first factor $\alpha_{ij}^0$ is a fitted rate constant which is independent of concentrations of $X \in$ \{ATP, ADP, P\} (see table \ref{tab:Kinesin_parameters}) or the external load force $F$ \cite{Liepelt}. The rate constants are depicted in table \ref{tab:rate_constants} and based on experiments by Visscher et al. \cite{Visscher1999}. If a transition $i\rightarrow j$ does not involve binding of ATP, ADP or P, the second factor $I_{ij}([X])$ is equal to one. Otherwise, it is linear dependent on the concentration of the involved reactant. For instance, ATP is bound during transition $1\rightarrow 2$. Hence, the transition rate $\alpha_{12}$ is linearly dependent on the concentration of ATP: $I_{12}([\text{ATP}])=[\text{ATP}]$.
The force dependence factor $\Phi_{ij}(F)$ is different for the mechanical transitions ($2\rightarrow 5$ and $5\rightarrow 2$) and all other chemical transitions.
The factors for the mechanical transitions are
\begin{equation}
\begin{aligned}
\Phi_{25}(F) &= \exp \left( - \theta \dfrac{Fl}{k_BT} \right) \\
\Phi_{52}(F) &= \exp \left( (1- \theta) \dfrac{Fl}{k_BT} \right),
\label{eq:mechanical_transitions}
\end{aligned}
\end{equation}
where $\theta$ denotes the dimensionless load distribution factor (see table \ref{tab:force_parameters}), $l$ the step size, $k_B$ the Boltzmann constant and $T$ the temperature (see table \ref{tab:Kinesin_parameters}).
For all chemical transitions, the force dependence factor
\begin{align}
\Phi_{ij}(F)=\dfrac{2}{1+e^{\chi_{ij}Fl/k_BT}} 
\label{eq:chemical_transitions}
\end{align}
is symmetric because the dimensionless force parameter $\chi_{ij}=\chi_{ji} \geqslant 0$ is symmetric.

\begin{table} \centering
\begin{tabular}{lll}
\hline \hline
Concentration            & $[P]$                                                                                     & $10^{-6}$\,M            \\
                         & $[ADP]$                                                                                   & $10^{-6}$\,M            \\
                         & $[ATP]$                                                                                   & $10^{-6}$\,M            \\ \hline  
Stepping size                 &     $l$                                                                                      & 8\,nm                   \\ \hline  
Thermal energy           &     $k_B T$                                                                                      & $4.1\cdot 10^{-21}$\,J  \\ \hline  
Fitted rate constants    & $\alpha_{25}^0$                                                                           & $3\cdot 10^5$\,$s^{-1}$ \\
                         & $\alpha_{52}^0$                                                                           & 0.24\,$s^{-1}$          \\
                         & $\alpha_{21}^0$                                                                           & 100\,$s^{-1}$           \\
                         & $\alpha_{54}^0=(\alpha_{52}^0/\alpha_{25}^0)^2\cdot \alpha_{21}^0$                      &                                        \\
                         & $\alpha_{56}^0=\alpha_{61}^0=\alpha_{23}^0=\alpha_{34}^0$  & 200\,$s^{-1}$           \\
                         & $\alpha_{65}^0=\alpha_{32}^0$                                                             & 0.09\,($\mu$Ms)$^{-1}$  \\
                         & $\alpha_{16}^0=\alpha_{43}^0$                                                             & 0.02\,($\mu$Ms)$^{-1}$  \\
                         & $\alpha_{12}^0=\alpha_{45}^0$                                                             & 1.8\,($\mu$Ms)$^{-1}$   \\ \hline  
Load distribution factor & $\theta$                                                                                  & 0.3                                    \\ \hline  
Force parameters         & $\chi_{12}=\chi_{45}$                                                                     & 0.25                                   \\
                         & $\chi_{23}=\chi_{56}$                                                                     & 0.05                                   \\
                         & $\chi_{34}=\chi_{61}$                                                                     & 0.05         \\                     \hline \hline    
\end{tabular}
\caption{Parameters determining the transition rates for the molecular motor kinesin in equation \eqref{eq:parameter_kinesin} \cite{Liepelt}. Fitted rate constants \cite{Visscher1999}, as given by equation \eqref{eq:parameter_kinesin}. Dimensionless load distribution factor $\theta$ which governs the mechanical transition rates, as given by equation \eqref{eq:mechanical_transitions}. Dimensionless force parameters which govern the force dependence of the chemical transition rates $\chi_{ij}=\chi_{ji}$ \cite{Liepelt, Visscher1999}, cf. equation \eqref{eq:chemical_transitions}. \label{tab:Kinesin_parameters} \label{tab:force_parameters} \label{tab:rate_constants}}
\end{table}

\subsection{Cycle fluxes in the kinesin network}
\begin{figure}
    \centering
    \includegraphics{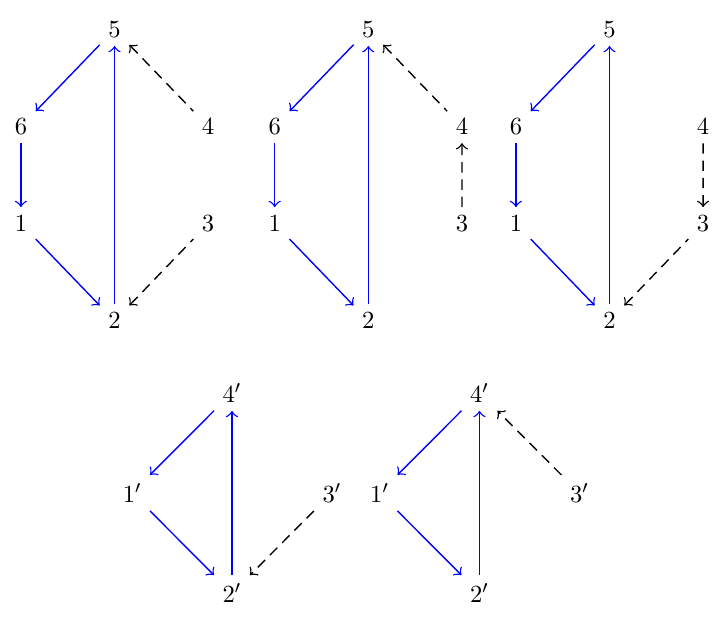}
    \caption{In the first row, the blue arrows contribute to the forward cycle $F$ in the positive ($+$) direction (counterclockwise). Its algebraic value is $\pi_{F+}=\alpha_{12}\alpha_{25}\alpha_{56}\alpha_{61}$. The dashed arrows flow into the forward cycle $F$. They correspond to the factor $S_C$ in equation \eqref{eq:4nodes3cycles-cylesfluxes}. In the second row, the diagrams for the forward cycle in plus direction are depicted for a coarse-grained network as shown in figure \ref{fig:Kinesin-network_CG}. States 1 and 6 in the original network are merged to state 1' in the coarse grained network. States 3 and 4 are merged to state 3'.  \label{fig:cycle_F+}}
\end{figure}

The cycle fluxes (net numbers of cycle completions per time) can be found from the extension of the diagram method described by Hill in \cite{hill}: A flux diagram for a cycle is the cycle itself plus a set of arrows flowing into it as depicted in figure \ref{fig:cycle_F+}. The algebraic value of a flux diagram for a cycle is the product of rate constants associated with the arrows multiplied by the contribution of the cycle (up to a normalisation constant).
The cycle fluxes in the Kinesin network can be calculated with
\footnotesize
\begin{equation}
\begin{aligned}
J_C &= \dfrac{\pi_C^+-\pi_C^-}{S}S_{C}=J_C^+-J_C^-\\
J_F &=(\alpha_{45}\alpha_{32}+\alpha_{34}\alpha_{45}+\alpha_{43}\alpha_{32})\dfrac{\alpha_{12}\alpha_{25}\alpha_{56}\alpha_{61}-\alpha_{16}\alpha_{65}\alpha_{52}\alpha_{21}}{S} \\
J_B &=(\alpha_{12}\alpha_{65}+\alpha_{61}\alpha_{12}+\alpha_{16}\alpha_{65})\dfrac{\alpha_{23}\alpha_{34}\alpha_{45}\alpha_{52}-\alpha_{25}\alpha_{54}\alpha_{43}\alpha_{32}}{S} \\
J_D&=\dfrac{\alpha_{12}\alpha_{23}\alpha_{34}\alpha_{45}\alpha_{56}\alpha_{61}-\alpha_{16}\alpha_{65}\alpha_{54}\alpha_{43}\alpha_{32}\alpha_{21}}{S},
\label{eq:6nodes3cycles-cylesfluxes}
\end{aligned}
\end{equation}
\normalsize
where the factor $S_C$ is the sum of all possible sets of remaining transitions, which flow into the cycle but do not form a cycle themselves. In figure \ref{fig:cycle_F+} these transitions are dashed. The factor $S$ is the sum of the weights of all spanning trees.
The cycle fluxes in the coarse-grained 4-state model can be calculated explicitly as
\begin{equation}
\begin{aligned}
\tilde{J}_C &= \dfrac{\tilde{\pi}_{C+}-\tilde{\pi}_{C-}}{\tilde{S}}\tilde{S}_{C}\\
\tilde{J}_F &=(\tilde{\alpha}_{34}+\tilde{\alpha}_{32})\dfrac{\tilde{\alpha}_{12}\tilde{\alpha}_{24}\tilde{\alpha}_{41}-\tilde{\alpha}_{14}\tilde{\alpha}_{42}\tilde{\alpha}_{21}}{\tilde{S}} \\
\tilde{J}_B &=(\tilde{\alpha}_{12}+\tilde{\alpha}_{14})\dfrac{\tilde{\alpha}_{23}\tilde{\alpha}_{34}\tilde{\alpha}_{42}-\tilde{\alpha}_{24}\tilde{\alpha}_{43}\tilde{\alpha}_{32}}{\tilde{S}} \\
\tilde{J}_D&=\dfrac{\tilde{\alpha}_{12}\tilde{\alpha}_{23}\tilde{\alpha}_{34}\tilde{\alpha}_{41}-\tilde{\alpha}_{14}\tilde{\alpha}_{43}\tilde{\alpha}_{32}\tilde{\alpha}_{21}}{\tilde{S}}.
\label{eq:4nodes3cycles-cylesfluxes}
\end{aligned}
\end{equation}

\subsection{Fluctuations in the kinesin network\label{sec:Fluctuations_explictly}}
Variances of steady-state observables can be explicitly calculated in terms of so-called one-way cycle fluxes \cite{CycleHill}. One-way cycle fluxes are not preserved by our coarse-graining method. Hence, variances are retained neither. 

\begin{figure}
    \centering
    \includegraphics{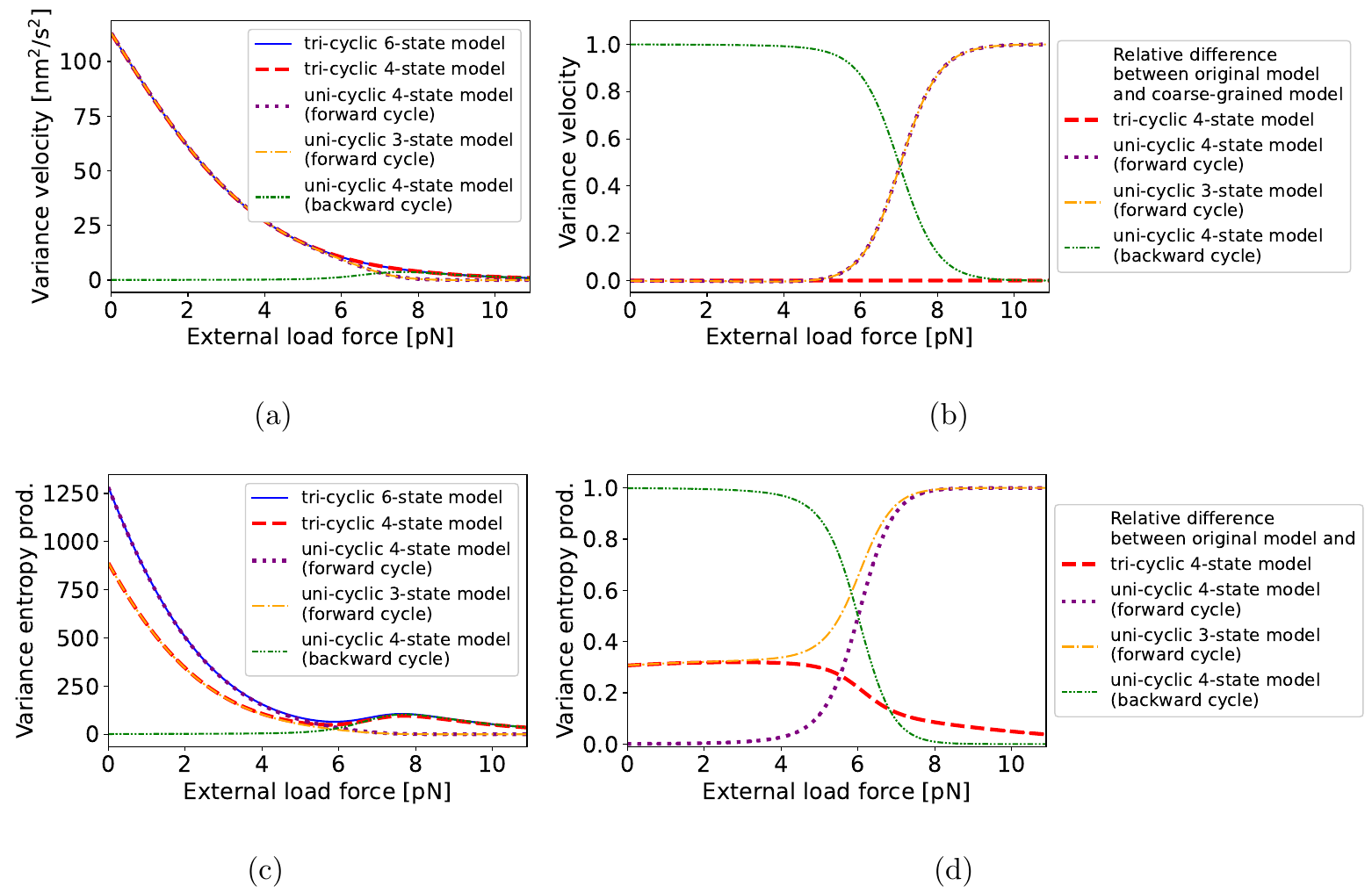}
    \caption{The variance of the velocity and the entropy production are plotted against the external load force acting on the motor for different kinesin models. 
    Furthermore, the relative differences between the variance of the velocity and the entropy production for coarse-grained models and the original 6-state model respectively are depicted in the second column. 
    The relative differences in the second column reveal large approximation errors by uni-cyclic systems that resembles the forward cycle for large external load forces.
    The motor has zero mean velocity for $F=7$\,pN. For no or small external load forces, uni-cyclic models approximate the variances for the velocity and the entropy production well with small deviations. While for larger load forces, a coarse-grained model is better, if it has the same network topology as the original model.
    Both quantities are calculated analytically as described in section \ref{sec:Fluctuations_explictly}. The variance was calculated for trajectories with length $\tau =1$\,s.\label{fig:distributions_variance}}
\end{figure}

The fluctuations of entropy production and velocity can be expressed in terms of fluctuations in the number of completed cycles. All transition fluxes can be written as a linear combination of cycle fluxes as shown in equation \eqref{eq:fluxes_Kinesin}. In equation \eqref{eq:4nodes3cycles-cylesfluxes}, it is demonstrated how to calculate the cycle fluxes in the kinesin network explicitly. The entropy production $P$ and velocity $v$ depend on transition fluxes. In order to calculate the variance of $P$ and $v$, we need to calculate the variance of transition fluxes, which are linear combinations of cycle fluxes. The latter can be calculated with Hill's diagram method \cite{hill}.
For a long time interval $\tau$, the net number $n_i$ of completed cycles of type $i\in \{F,B,D\}$ can be treated as an independent random variable which has a Gaussian distribution with mean and variance
\begin{equation}
\begin{aligned}
\bar{n}_i (\tau ) &=J_i^{net} \tau \\
\sigma^2_i (\tau) &=(J_{i+}+J_{i-})\tau ,
\label{eq:hill_variance}
\end{aligned}
\end{equation}
as shown in \cite{hill2012free}, where $J_{i\pm}$ denotes the cycle flux in positive or negative direction respectively. To avoid confusion, the net-transition flux is denoted as $J_i^{net}$. But if not stated otherwise all fluxes are net fluxes.
The net flux through edge (2,5) can be expressed in terms of cycle fluxes as proposed in equation \eqref{eq:fluxes_Kinesin}, such that
\begin{align}
J^{net}_{25}=J_F^{net}-J_B^{net}.
\end{align}
The net numbers of completed forward (F) and backward (B) cycles (depicted in figure \ref{fig:Cycle_Decomposition}) are independent random variables \cite{hill2012free} for long time intervals $\tau$. Thus, the variance adds up, as the net number of transitions $n_{25}$ is a sum of independent random variables, which have zero covariance:
\begin{equation}
\begin{aligned}
n_{25}&=n_F-n_B \\
\bar{n}_{25}&=\bar{n}_F - \bar{n}_B \\
\sigma_{25}^2 &=\sigma_{F}^2+\sigma_{B}^2.
\end{aligned}
\end{equation}
The velocity thus has mean and variance
\begin{equation}
\begin{aligned}
\bar{\dfrac{v}{l}}&= J_{25}^{net} =\dfrac{\bar{n}_{25}}{\tau} \\
\sigma^2_{v/l} &= \dfrac{\sigma^2_{n_{25}}}{\tau^2  }= \dfrac{J_{F+}+J_{F-}+ J_{B+}+J_{B-} }{\tau}. 
\label{eq:Mean_Var_velocity}
\end{aligned}
\end{equation}
An explicit expression for the one-way cycle fluxes can be found in equation \eqref{eq:4nodes3cycles-cylesfluxes}.
If a coarse-graining iteration preserves the cycle topology (no cycle is lost due to a merged pair of states), the single-cycle fluxes, in general, are not retained. But the sum of net-cycle fluxes (which corresponds to a transition flux) remains unchanged. Thus, the mean of the velocity is retained in all coarse-graining iterations. 
The variance is not preserved because one-way cycle fluxes are not retained by our coarse-graining approach. By contrast, one-way transition fluxes are preserved but cannot be expressed in terms of one-way cycle fluxes.
The entropy production
\begin{equation}
\begin{aligned}
P&= J_{12} \Delta S_F +J_{23} \Delta S_B \\
&= (J_F+J_D)\Delta S_F + (J_B + J_D) \Delta S_B \\
&= \dfrac{n_F+n_D}{\tau} \Delta S_F + \dfrac{n_B+n_D}{\tau} \Delta S_B \\
\end{aligned}
\end{equation}
has variance
\begin{equation}
\begin{aligned}
\sigma^2_P &= (\sigma^2_{n_F}+\sigma^2_{n_D})\left( \dfrac{\Delta S_F}{\tau} \right)^2 + (\sigma^2_{n_B}+\sigma^2_{n_D}) \left( \dfrac{\Delta S_B}{\tau} \right)^2 \\
&= \dfrac{J_{F+}+J_{F-}+J_{D+}+J_{D-}}{\tau} (\Delta S_F)^2 \\ &+
 \dfrac{J_{B+}+J_{B-}+J_{D+}+J_{D-}}{\tau} (\Delta S_B)^2,
\end{aligned}
\end{equation}
where we expressed the chord fluxes in terms of cycle fluxes as given by equation \eqref{eq:fluxes_Kinesin} and used that for long trajectories, the numbers of completed cycles are independent Gaussian distributed random variables as given by equation \eqref{eq:hill_variance} \cite{CycleHill}.
To summarize, the mean of the velocity of the kinesin motor is preserved by the coarse-graining procedure, whereas the mean of the entropy production and variances for entropy production and velocity are not preserved.
Figure \ref{fig:distributions_variance} shows the variance of the velocity and the entropy production as a function of the external load force acting on the motor for tri-cyclic and uni-cyclic kinesin models.
\newpage
\bibliography{literatur}
\end{document}